# Electronic Structure, mass fluctuation, and Localized Bond Properties of two-dimensional double-layer transition metal chalcogenide $MX_2$ (M = Mo, W; X = S, Se, Te) Calculated Based on Density Functional Theory and BBC model

Yaorui Tan and Maolin Bo*
Key Laboratory of Extraordinary Bond Engineering and Advanced Materials Technology (EBEAM) of Chongqing, Yangtze Normal University, Chongqing 408100, China
*Author to whom any correspondence should be addressed.
(E-mail addresses: bmlwd@yznu.edu.cn)

**Abstract**

This study systematically investigates the electronic structure and bonding properties of two-dimensional bilayer transition metal chalcogenides $MX_2$ (M = Mo, W; X = S, Se, Te) using density functional theory calculations. By analyzing band gaps, deformation bond energies, and non-Hermitian bonding characteristics across various $MX_2$ compounds, we comprehensively examine their electronic properties and chemical bonding behavior. The results reveal that charge transfer plays a crucial role in electron mass fluctuations, with topological geometric analysis further confirming the impact of mass variations on atomic bonding and electronic states. These findings provide a theoretical foundation for advancing the application of these materials.

## 1. Introduction

With the widespread application of two-dimensional materials in electronics, optics, and catalysis, transition metal chalcogenides $MX_2$ (M = Mo, W; X = S, Se, Te) have emerged as a focal point of research due to their distinctive physical and chemical properties. For instance, transition metal chalcogenide heterojunctions, used as electrode materials in lithium-ion batteries [1, 2], significantly enhance structural stability and conductivity, leading to improved specific capacity and cycling performance. Additionally, materials such as molybdenum disulfide ($MoS_2$), which possess relatively large interlayer spacing, facilitate lithium-ion intercalation and deintercalation, demonstrating exceptional performance in battery applications[3-5].

Beyond energy storage, transition metal chalcogenide heterojunctions exhibit great potential in photocatalysis[6-8], particularly in water splitting for hydrogen and oxygen generation, enabling the efficient conversion of solar energy into chemical energy. Constructing heterojunctions enhances the separation efficiency of photogenerated charge carriers [9, 10], thereby improving photocatalytic activity and offering innovative solutions for clean energy production. Moreover, these materials serve as effective light-absorbing layers in solar cells[11], where their tunable bandgap and strong light absorption capabilities optimize energy band structures, ultimately enhancing photovoltaic conversion efficiency. By selecting different transition metals and chalcogen elements and adjusting atomic layer stacking, the bandgap of these materials can be precisely controlled, providing valuable theoretical guidance for the design of diverse optoelectronic devices[12-15]. Consequently, transition metal chalcogenide heterojunctions hold significant promise not only in energy storage and optoelectronics but also in driving technological advancements in related fields.

This study employs density functional theory (DFT) and the Binding Energy and Bond Charge (BBC) model[16] to analyze the band structure, density of states, and atomic bonding in two-dimensional bilayer $MX_2$ compounds. The band gaps of the heterostructures $MoS_2/WSe_2$ and $WSe_2/MoTe_2$ [17-20]were determined to be 0.934 eV and 0 eV, respectively, while the band gaps of bilayer $WS_2$, $WSe_2$, and $WTe_2$ [21-23] were 1.838 eV, 1.496 eV, and 0.917 eV. These results highlight the substantial variability in band gaps, underscoring the ability of interlayer heterostructures to modulate the bandgap of two-dimensional transition metal dichalcogenides. Furthermore, calculations of electron transfer and atomic bonding reveal that charge transfer is intricately linked to electronic mass fluctuations, demonstrating that these variations significantly influence atomic bonding and the overall electronic structure.

## 2. Methods

### 2.1 DFT calculations

The electronic properties and structural relaxation of $MoS_2/WSe_2$, $WSe_2/MoTe_2$, $WS_2$, $WSe_2$, and $WTe_2$ were simulated using the Cambridge Sequential Total Energy Package (CASTEP). This computational package employs DFT based on the plane-wave pseudopotential approach and integrates the Perdew–Burke–Ernzerhof (PBE) functional to handle electron exchange and its associated potential energy. The TS scheme was applied for DFT-D dispersion correction to account for long-range van der Waals (vdW) interactions. For the band structure calculations, the HSE06 functional [24] was utilized.

The simulations examined the energy, structure, bonding, and electronic properties of the aforementioned compounds: $MoS_2/WSe_2$, $WSe_2/MoTe_2$, $WS_2$, $WSe_2$, and $WTe_2$. The cut-off energies, band gaps, and k-point grids used in this study are summarized in Table 1. All structures were fully optimized without symmetry constraints until the forces were reduced to below 0.03 eV/Å, and the energy tolerance decreased below $1\times10^{-5}$ eV the specified threshold per atom. For self-consistent field calculations, convergence threshold $1.0\times10^{-6}$ eV per atom was employed for $MoS_2/WSe_2$, $WSe_2/MoTe_2$, $WS_2$, $WSe_2$, and

Additionally, we selected structures with lattice strain below 5% using QuantumWise, and the lattice parameters for each structure, obtained from QuantumWise, are provided in Table 1.

**Table 1** Cut-off energies, bandgaps, and *k*-points of $MoS_2/WSe_2$, $WSe_2/MoTe_2$, $WS_2$, $WSe_2$ and $WTe_2$

| | lattice strain | Cut-off energy | *k*-point | Band gap (HSE06) |
|---|---|---|---|---|
| $MoS_2/WSe_2$ | 2.73 % | 650.0 eV | 15×15×2 | 0.934 eV |
| $WSe_2/MoTe_2$ | 4.32% | 450.0 eV | 14×14×2 | 0.334 eV |
| $WS_2$ | 0% | 650.0 eV | 14×14×2 | 1.838 eV |
| $WSe_2$ | 0% | 450.0 eV | 14×14×2 | 1.496 eV |
| $WTe_2$ | 0% | 450.0 eV | 14×14×2 | 0.917 eV |

### 2.2. BBC model

The BBC model[16] comprises the binding energy model (BC), bond charge model (BC), and the Hamiltonian of the BB model **(see supplemental material)**.

$$\begin{cases} H = \xi(0)\sum_{I}\hat{n}_I - J\sum_{I}\sum_{\rho}C_I^+C_{I+\rho} = \left(E_n^a - \gamma^m A_n\right)\sum_{I}\hat{n}_I - J\sum_{I}\sum_{\rho}C_I^+C_{I+\rho} \\ E_n(k) = E_n^a - \gamma^m A_n - J_n\sum_{\rho}e^{i\vec{k}\vec{\rho}} \end{cases}$$

(1)

$\hat{n}_l$ represents the electron number operator in the wanier representation. So, it $\sum_{l}\hat{n}_l = C_l^+C_l$ can be obtained. $A_n$ represents the energy shift, $E_n^a$ is the atomic energy level, and $J$ is called the overlapping integral.

$$\begin{cases} A_n = -\int \phi^*\left(\vec{r}-\vec{l}\right)\left[V\left(\vec{r}\right) - v_a\left(\vec{r}-\vec{l}\right)\right]\phi\left(\vec{r}-\vec{l}\right)\mathrm{d}r \\ E_n^a = \int \phi^*\left(\vec{r}-\vec{l}\right)\left[-\frac{h^2}{2m}\nabla^2 - v_a\left(\vec{r}-\vec{l}\right)\right]\phi\left(\vec{r}-\vec{l}\right)\mathrm{d}r \\ J_n = -\int \phi^*\left(\vec{r}\right)\left[V\left(\vec{r}\right) - v_a\left(\vec{r}-\vec{l}\right)\right]\phi\left(\vec{r}-\vec{\rho}\right)\mathrm{d}r \end{cases}$$

（2）

$\xi(0)$ is the local orbital electron energy. Therefore, there exists $\xi(0)=\left(E_n^a - A_n\right)$. $E_n^a$ caused by the potential of $N$-1 atoms outside the lattice point $l$ in the lattice $V(r) = -\sum_{i,j,l_j\neq 0}\frac{1}{4\pi\varepsilon_0}\frac{Z'e^2}{\left|\vec{r}_i - \vec{l}_j\right|}$ is the potential and $v_a\left(\vec{r}-\vec{l}\right)$ is the atomic potential.

The binding energy (BE) shifts of the bulk and surface atoms can be expressed as:

$$\Delta E_v\left(B\right) = A_n,\ \ \Delta E_v\left(i\right) = \gamma^m A_n$$

(3)

For chemical bonding, **Eq. 2** can be written as:

$$\gamma^m = \frac{E_v(i) - E_v(0)}{E_v(B) - E_v(0)} = \left(\frac{E_v(x) - E_v(0)}{E_v(B) - E_v(0)}\right)^m \approx \left(\frac{Z_x d_b}{Z_b d_x}\right)^m = \left(\frac{Z_b - \sigma_i'}{Z_b}\right)^m \left(\frac{d_x}{d_b}\right)^{-m} \approx \left(\frac{d_x}{d_b}\right)^{-m}$$

(4)

$\sigma'$ is very small; therefore $\left(Z_b - \sigma'\right)/Z_b \approx 1$ , For metal $m$ =

1, $\left(\frac{Z_b-\sigma_i'}{Z_b}\right)^m\left(\frac{d_x}{d_b}\right)^{-m}=\left(\frac{d_x}{d_b}\right)^{-m'}$ obtain $m'=m\left(1-\frac{\ln\frac{Z_b-\sigma_i'}{Z_b}}{\ln\left(\frac{d_x}{d_b}\right)}\right)$. For example, we calculate the energy level shift of Mo 3$d$, $E_v(B)$ = 227.569 eV, $E_v(x)$ = 227.957 eV, $\frac{d_x}{d_b}$ =0.875; We use $E_i=-13.6\frac{(Z-\sigma_i)^2}{n^2}$, considering the influence of angular momentum $E_i(x)=-13.6\frac{(Z-\sigma_i)^2}{(n+0.7*l)^2}$, $Z$ = 42, $n$ = 3, $l$ = 2. For the orbital energy of Mo 3d: $E_{3d}(x)=-13.6\frac{(42-\sigma_i)^2}{(3+0.7*2)^2}$, $\sigma_B$ = 24.001 eV, $\sigma_x$ = 23.986 eV, $\sigma_i'=0.015$, $m'=0.995$ $\approx 1$. Therefore $m\approx m'$.

The Hamiltonian of a system in the BC model is expressed as **(see supplemental material)**:

$$H=\sum_{k\sigma}\frac{\hbar^2k^2}{2m}a_{k\sigma}^{\dagger}a_{k\sigma}+\frac{e_1^2}{2V}\sum_q{}^{*}\sum_{\vec{k}\sigma}\sum_{\vec{K'}\lambda}\frac{4\pi}{q^2}a_{\vec{k}+\vec{q},\sigma}^{\dagger}a_{\vec{K'}-\vec{q},\lambda}^{\dagger}a_{\vec{K'}\lambda}a_{\vec{k}\sigma}\tag{5}$$

The electron-interaction terms for density fluctuations is expressed as:

$$\begin{aligned}\Delta H&=H_{e1}-H_{e2}\\&=\frac{e_1^2}{2V}\sum_q{}^{*}\sum_{\vec{k}\sigma}\sum_{\vec{K'}\lambda}\frac{4\pi}{q_1^2}a_{\vec{k}+\vec{q},\sigma}^{\dagger}a_{\vec{K'}-\vec{q},\lambda}^{\dagger}a_{\vec{K'}\lambda}a_{\vec{k}\sigma}-\frac{e_1^2}{2V}\sum_q{}^{*}\sum_{\vec{k}\sigma}\sum_{\vec{K'}\lambda}\frac{4\pi}{q_2^2}a_{\vec{k}+\vec{q},\sigma}^{\dagger}a_{\vec{K'}-\vec{q},\lambda}^{\dagger}a_{\vec{K'}\lambda}a_{\vec{k}\sigma}\\&=\frac{1}{4\pi\varepsilon_0}\sum_{i=1}^{N}\sum_{\substack{j=1\\i\neq j}}^{N}\frac{e^2}{2\left|\vec{r_i}-\vec{r_j}\right|}-\frac{1}{4\pi\varepsilon_0}\sum_{i=1}^{N}\sum_{\substack{j=1\\i\neq j}}^{N}\frac{e^2}{2\left|\vec{r_i}'-\vec{r_j}'\right|}\end{aligned}$$

According to charge conservation, the following formula describes charge density fluctuations:

$$\begin{aligned}\delta\hat{V}_{12}&=V_{e1}-V_{e2}\\&=\sum_{i,j}\frac{1}{4\pi\varepsilon_0}\frac{e^2}{2\left|\vec{r_i}-\vec{r_j}\right|}\int d^3\vec{r_i}\int d^3\vec{r_j}\rho(\vec{r_i})\rho(\vec{r_j})-\sum_{i,j}\frac{1}{4\pi\varepsilon_0}\frac{e^2}{2\left|\vec{r_i}'-\vec{r_j}'\right|}\int d^3\vec{r_i}'\int d^3\vec{r_j}'\rho(\vec{r_i}')\rho(\vec{r_j}')\\&=\sum_{i,j}\frac{1}{4\pi\varepsilon_0}\frac{e^2}{2\left|\vec{r}-\vec{r'}\right|}\int d^3r\int d^3r'\delta\rho(\vec{r})\delta\rho(\vec{r'})\end{aligned}\tag{6}$$

The deformation bond energy $\delta V_{bc}$ is represented by **Eq. 7**:

$$\delta V_{bc} = \frac{1}{4\pi\varepsilon_0}\frac{e^2}{2\left|\vec{r}-\vec{r'}\right|}\int d^3 r \int d^3 r' \delta\rho\left(\vec{r}\right)\delta\rho\left(\vec{r'}\right) \tag{7}$$

Also, $\delta\rho$ satisfies the conditions outlined in **Eq.8**:

$$\left(\delta\rho_{Hole\text{-}electron} \le \delta\rho_{Antibonding\text{-}electron} < \delta\rho_{No\ charge\ tranfer} = 0 < \delta\rho_{Nonbonding\text{-}electron} \le \delta\rho_{Bonding\text{-}electron}\right) \tag{8}$$

For atomic (strong) bonding states:

$$\delta\rho_{Hole\text{-}electron}\left(\vec{r}\right)\delta\rho_{Bonding\text{-}electron}\left(\vec{r}'\right) < 0(Strong\ Bonding) \tag{9}$$

For atomic nonbonding or weak-bonding states:

$$\begin{cases} \delta\rho_{Hole\text{-}electron}\left(\vec{r}\right)\delta\rho_{Nonbonding\text{-}electron}\left(\vec{r}'\right) < 0(Nonbonding\ or\ Weak\ Bonding) \\ \delta\rho_{Antibonding\text{-}electron}\left(\vec{r}\right)\delta\rho_{Bonding\text{-}electron}\left(\vec{r}'\right) < 0(Nonbonding\ or\ Weak\ Bonding) \\ \delta\rho_{Antibonding\text{-}electron}\left(\vec{r}\right)\delta\rho_{Nonbonding\text{-}electron}\left(\vec{r}'\right) < 0(Nonbonding) \end{cases} \tag{10}$$

For atomic antibonding states:

$$\begin{cases} \delta\rho_{Nonbonding\text{-}electron}\left(\vec{r}\right)\delta\rho_{Bonding\text{-}electron}\left(\vec{r}'\right) > 0(Antibonding) \\ \delta\rho_{Hole\text{-}electron}\left(\vec{r}\right)\delta\rho_{Antibonding\text{-}electron}\left(\vec{r}'\right) > 0(Antibonding) \\ \delta\rho_{Hole\text{-}electron}\left(\vec{r}\right)\delta\rho_{Hole\text{-}electron}\left(\vec{r}'\right) > 0(Antibonding) \\ \delta\rho_{Antibonding\text{-}electron}\left(\vec{r}\right)\delta\rho_{Antibonding\text{-}electron}\left(\vec{r}'\right) > 0(Antibonding) \\ \delta\rho_{Nonbonding\text{-}electron}\left(\vec{r}\right)\delta\rho_{Nonbonding\text{-}electron}\left(\vec{r}'\right) > 0(Antibonding) \\ \delta\rho_{Bonding\text{-}electron}\left(\vec{r}\right)\delta\rho_{Bonding\text{-}electron}\left(\vec{r}'\right) > 0(Antibonding) \end{cases} \tag{11}$$

The formation of chemical bonds is related to fluctuations in electron density $\delta\rho$.

### 2.3 Riemann sphere and Möbius transformation

In Euclidean space $R^3$, the complex number set $C$ is represented as the $XY$-plane[24]. A sphere $S$ is defined with center at (0,0,1/2) and radius1/2, with the north pole $N$ = (0,0,1) as its vertex. The point $W$ = (X, Y, Z) lies on the surface of sphere $S$, and the line connecting $N$ and $W$ intersects the $XY$-plane at the projection point $w = x + iy$. Conversely, a line from the point $w$= ($x$, $y$, 0) in the complex plane to $N$ intersects the sphere at a unique point $W$. As shown in **Fig. 1**.

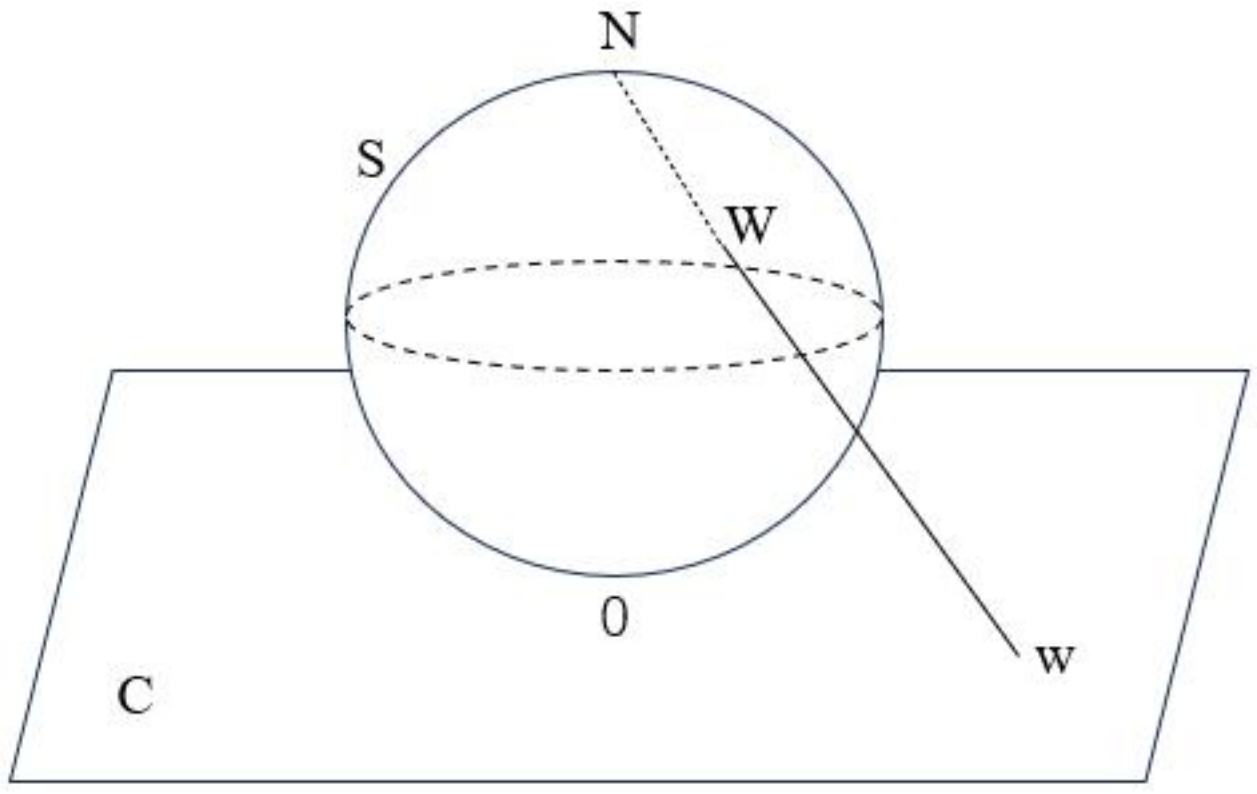

**Fig. 1** Riemann sphere $S$ and projection

This geometric interpretation provides an objective representation of a point on a perforated spherical surface $S$ and a complex plane. The analytical description of this objective representation is as follows.

$$x=\frac{X}{1-Z},\quad y=\frac{Y}{1-Z} \tag{12}$$

Substituting $W$ with $w$:

$$Y=\frac{y}{x^2+y^2+1},\quad X=\frac{x}{x^2+y^2+1},\quad Z=\frac{x^2+y^2}{x^2+y^2+1} \tag{13}$$

The complex plane can cover the sphere, and a rational function defined on the extended complex plane can be interpreted as a mapping of the sphere onto itself, transforming a given pole into the north pole N. The Riemann sphere provides a powerful geometric framework for understanding the relationship between the complex number set and rational functions.

The Möbius transformation[25] maps a point $z$ on the complex plane, expressed as $M(z)=\frac{az+b}{cz+d}$ with complex numbers $a$, $b$, $c$, and $d$ (where $ad-bc\neq 0$ ). It is regarded as a fractional linear function that uniquely maps each point on the complex plane. Each Möbius transformation $M(z)$ corresponds to a $(2\times 2)$ matrix, $[M]$, with complex elements:

$$M(z)=\frac{az+b}{cz+d}\rightarrow[M]=\begin{bmatrix} a & b \\ c & d \end{bmatrix}$$

(14)

To visualize the Möbius transformation $M(z)=\dfrac{az+b}{cz+d}$, we assume it has two fixed points $\xi_{-}$ and consider it a mapping of itself $z\mapsto\omega=M(z)$, as shown in **Fig. 2a**.

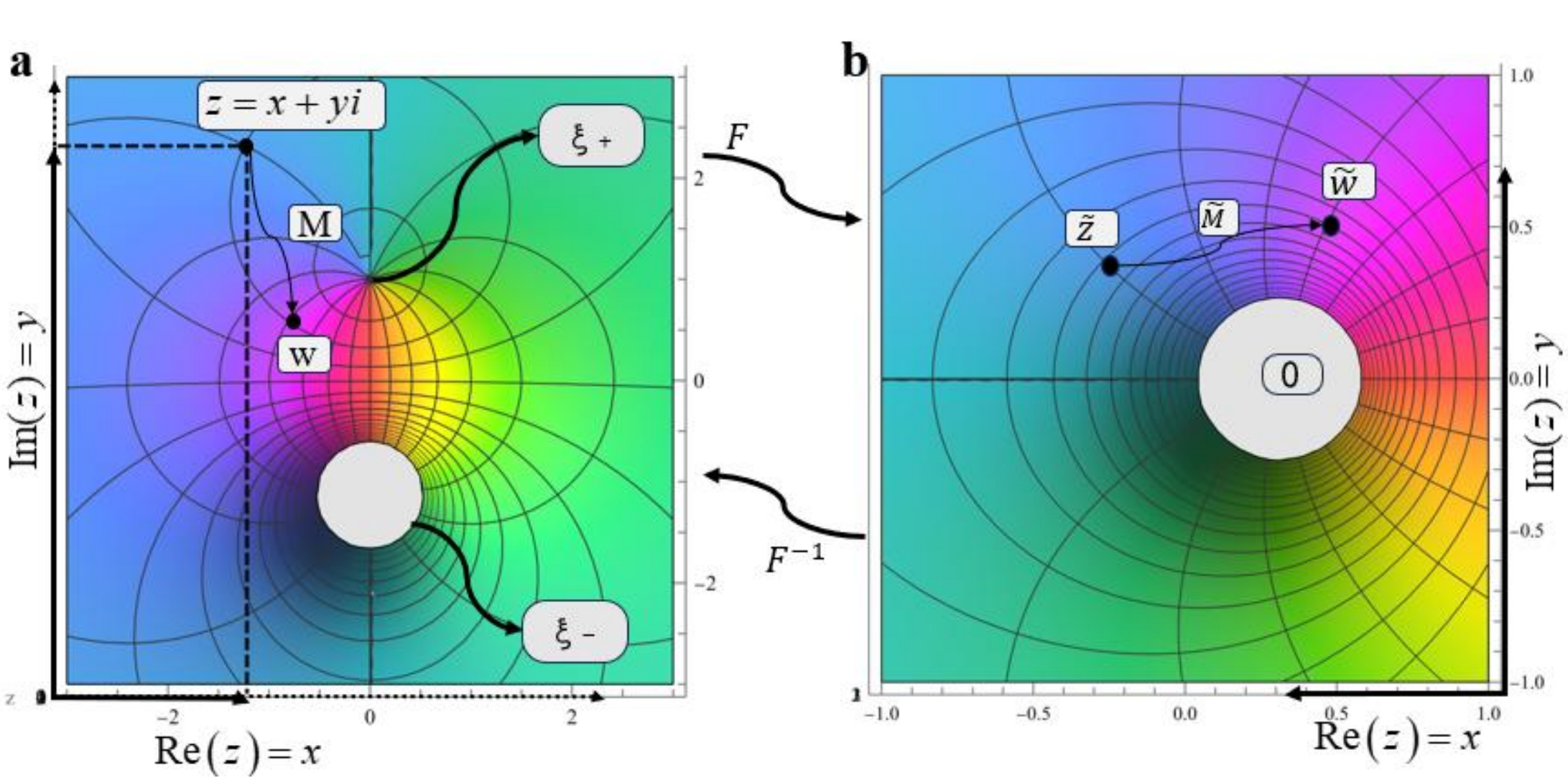

**Fig. 2** Möbius transformation diagram

In the simplest Möbius transformation (see **Fig. 2a**), $F(z)=\dfrac{z-\xi_+}{z-\xi_-}$ the transformation maps one fixed point (set to $\xi_+$) to zero and the other fixed point ($\xi_-$) to $\infty$ infinity. **Fig. 2b** illustrates the image on the left under this Möbius transformation.

**2.4 Lorenz transformation**

Einstein discovered that there exists a spacetime interval $\Gamma$ [26] between an observer and an event, such that the values of the spacetime interval seen by two observers are consistent. This spacetime interval $\Gamma$ is defined by its square:

$$\Gamma^2=T^2-\left(X^2+Y^2+Z^2\right)=\tilde{T}^2-\left(\tilde{X}^2+\tilde{Y}^2+\tilde{Z}^2\right)$$

(15)

Minkowski recognized that this interval was a proper extension of the concept of distance, making it suitable for use in spacetime. The equidistant transformation and the symmetry of spacetime ensure the invariance of this interval. However, it differs significantly from conventional distance: the square of the interval between different events can be zero or even negative. If two observers focus on the same photon, what they see is consistent, with each event along the photon's trajectory in spacetime having $\Gamma = 0$. This four-dimensional vector, which points to a specific light ray and has its "length" vanish, is called a zero vector.

The Lorentz transformation L is a linear transformation in spacetime, essentially a (4×4) matrix that maps an observer's observation $\left(T, X, Y, Z\right)$ of an event to another observer's observation $\left(\tilde{T}, \tilde{X}, \tilde{Y}, \tilde{Z}\right)$ of the same event. In other words, L maintains a constant spatiotemporal distance, and the observations of two observers on the spatiotemporal interval of the same event are consistent.

Consider an event $O$ that generates an electric spark, emitting photons that propagate uniformly in all directions, forming a spherical surface centered at the origin. As time progresses, the radius of the spherical surface increases. At the time $T = 1$, the photon forms a sphere with a radius of 1, which is now represented by a Riemann sphere. Thus, it can be considered that the sphere composed of these photons consists of points marked with spacetime coordinates $\left(1, X, Y, Z\right)$, which can also be represented by complex numbers.

The projection formula $z = \left(z_1 / z_2\right)$ onto the polar plane of the ball is given by:

$$X = \frac{\varsigma_1\overline{\varsigma}_2 + \varsigma_2\overline{\varsigma}_1}{\left|\varsigma_1\right|^2 + \left|\varsigma_2\right|^2}, Y = \frac{\varsigma_1\overline{\varsigma}_2 - \varsigma_2\overline{\varsigma}_1}{i\left(\left|\varsigma_1\right|^2 + \left|\varsigma_2\right|^2\right)}, Z = \frac{\left|\varsigma_1\right|^2 - \left|\varsigma_2\right|^2}{\left|\varsigma_1\right|^2 + \left|\varsigma_2\right|^2} \tag{16}$$

However, each beam of light is equivalent to any zero vector in the direction of that beam of light. If we let $T = 1$, we can choose a scalar factor $\left|\varsigma_1\right|^2 + \left|\varsigma_2\right|^2$ multiplied by the expression, which eliminates the denominator ($T = \left|\varsigma_1\right|^2 + \left|\varsigma_2\right|^2$). This new zero vector $\left(T, X, Y, Z\right)$ (along the original spatiotemporal direction) can be represented as

$$\begin{aligned} T &= \left|\varsigma_1\right|^2 + \left|\varsigma_2\right|^2, \\ X &= \varsigma_1\overline{\varsigma}_2 + \varsigma_2\overline{\varsigma}_1, \\ Y &= -i\left(\varsigma_1\overline{\varsigma}_2 - \varsigma_2\overline{\varsigma}_1\right), \\ Z &= \left|\varsigma_1\right|^2 - \left|\varsigma_2\right|^2 \end{aligned} \tag{17}$$

It is straightforward to convert these formulas into the following form:

$$\begin{pmatrix} T+Z & X+iY \\ X-iY & T-Z \end{pmatrix} = 2\begin{pmatrix} \varsigma_1\bar{\varsigma}_1 & \varsigma_1\bar{\varsigma}_2 \\ \varsigma_2\bar{\varsigma}_1 & \varsigma_2\bar{\varsigma}_2 \end{pmatrix} = 2\begin{pmatrix} \varsigma_1 \\ \varsigma_2 \end{pmatrix}\begin{pmatrix} \bar{\varsigma}_1 & \bar{\varsigma}_2 \end{pmatrix} \tag{18}$$

which simplifies to:

$$\begin{pmatrix} T+Z & X+iY \\ X-iY & T-Z \end{pmatrix} = 2\begin{pmatrix} \varsigma_1 \\ \varsigma_2 \end{pmatrix}\begin{pmatrix} \varsigma_1 \\ \varsigma_2 \end{pmatrix}^* \tag{19}$$

Here, the symbol “*” represents conjugate transposition. The spatiotemporal interval can now be succinctly represented as the determinant of this matrix:

$$\Gamma^2 = T^2 - \left(X^2 + Y^2 + Z^2\right) = \det\begin{pmatrix} T+Z & X+iY \\ X-iY & T-Z \end{pmatrix} \tag{20}$$

It is evident that this enlarged spatiotemporal vector remains a zero vector, as expected. the complex projection of the spherical polar plane has been employed to illustrate the effect of the Möbius transformation $z \mapsto \tilde{z} = M\left(z\right)$ on this beam of light on the complex plane $\mathbb{C}$, as follows:

$$\begin{bmatrix} \varsigma_1 \\ \varsigma_2 \end{bmatrix} \mapsto \begin{bmatrix} \tilde{\varsigma}_1 \\ \tilde{\varsigma}_2 \end{bmatrix} = \left[M\right]\begin{bmatrix} \varsigma_1 \\ \varsigma_2 \end{bmatrix} \tag{21}$$

This results in a linear transformation of the zero vector generated by the light trajectory:

$$\begin{pmatrix} T+Z & X+iY \\ X-iY & T-Z \end{pmatrix} \mapsto \begin{pmatrix} \tilde{T}+\tilde{Z} & \tilde{X}+i\tilde{Y} \\ \tilde{X}-i\tilde{Y} & \tilde{T}-\tilde{Z} \end{pmatrix} = \left[M\right]\begin{pmatrix} T+Z & X+iY \\ X-iY & T-Z \end{pmatrix}\left[M\right]^* \tag{22}$$

Finally, consider the application of this linear transformation to all spatiotemporal vectors (not just zero vectors). Since $\det\left[M\right] = 1 = \det\left[M\right]^*$, it can be concluded that this linear transformation preserves the spatiotemporal interval:

$$\tilde{\Gamma}^2 = \det\left\{\left[M\right]\begin{pmatrix} T+Z & X+iY \\ X-iY & T-Z \end{pmatrix}\left[M\right]^*\right\} = \Gamma^2 \tag{23}$$

Each Möbius transformation on a complex plan $\mathbb{C}$ corresponds to a unique Lorentz transformation in spacetime. Conversely, it can be proven that each Lorentz transformation has a corresponding unique Möbius transformation. Thus, a one-to-one correspondence exists between the Möbius transformation and the Lorentz transformation, demonstrating the invariance of spatiotemporal geometry across different reference frames.

2.5 Fluctuation of electron mass

An electron and a positron can interact through the exchange of a photon, which can, in turn, transform into an electron-positron pair [28]. As shown in **Fig. 3**, the shaded region represents the vacuum polarization tensor, denoted as $i\prod_{\mu\nu}(q)$. Based on an infinite series of Feynman diagrams, as illustrated in **Fig. 4**, an approximation using the lowest-order diagram in **Fig. 3** with the vacuum polarization tensor results in the diagram shown in **Fig. 5.**

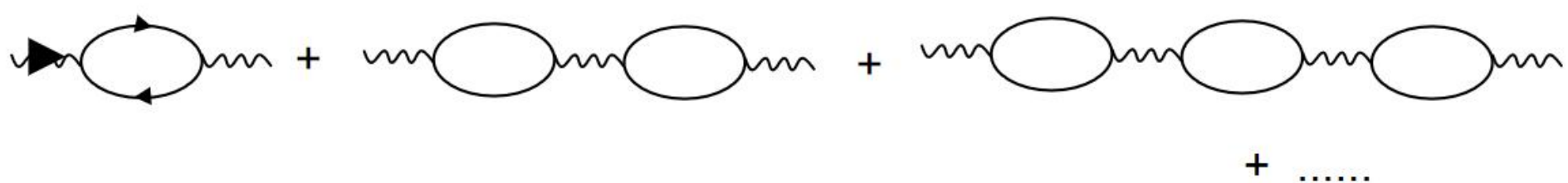

**Fig. 3** the process of mutual annihilation

**Fig. 4** physical photon propagation

**Fig. 5** infinite Feynman diagrams

The Lagrangian for this transformation $A \to (1/e)A$ is given by:

$$L=\overline{\psi}\left[i\gamma^{\mu}\left(\partial_{\mu}-iA_{\mu}\right)-m\right]\psi-\frac{1}{4e^{2}}F_{\mu\nu}F^{\mu\nu}$$

(24)

At $\psi \to e^{i\alpha}\psi, A_\mu \to A_\mu + \partial_\mu \alpha$ this point,the photon propagator taking the inverse of $\frac{1}{4e^2}F_{\mu\nu}F^{\mu\nu}$ is proportional to $e^2$ :

$$iD_{\mu\nu}(q)=\frac{-ie^2}{q^2}\left(g_{\mu\nu}-(1-\xi)\frac{q_\mu q_\nu}{q^2}\right)$$

(25)

According to the combination of gauge-invariant diagrams and Lorentz invariance:

$$\begin{cases} q_\mu \prod_{\mu\nu}(q)=0 \\ \prod_{\mu\nu}(q)=\left(q_\mu q_\nu - g_{\mu\nu}q^2\right)\prod\left(q^2\right) \end{cases}$$

(26)

The physical photon propagation in **Fig. 4** is represented by the following geometric series:

$$\begin{aligned} iD^P_{\mu\nu}(q) &= iD_{\mu\nu}(q)+iD_{\mu\lambda}(q)i\prod^{\lambda\rho}(q)iD_{\rho\nu}(q) \\ &\quad + iD_{\mu\lambda}(q)i\prod^{\lambda\rho}(q)iD_{\rho\sigma}(q)i\prod^{\sigma\kappa}(q)iD_{\kappa\nu}(q)+\cdots \\ &= \frac{-ie^2}{q^2}g_{\mu\nu}\frac{1}{1+e^2\prod\left(q^2\right)}+q_\mu q_\nu \end{aligned}$$

(27)

$(1-\xi)\frac{q_\mu q_\nu}{q^2}$ in part $iD_{\mu\nu}(q)=\frac{-ie^2}{q^2}\left(g_{\mu\nu}-(1-\xi)\frac{q_\mu q_\nu}{q^2}\right)$ encounters $\prod_{\lambda\rho}(q)$ as 0, in **Eq. 27**, and the parameter $\xi$ appears only in the $q_\mu q_\nu$ term and not in the physical amplitude, in **Eq. 26**.

The residue of the pole $iD^P_{\mu\nu}(q)$ is the square of the heavy positive charge, also known as physics:

$$e_R^2=e^2\frac{1}{1+e^2\prod(0)}$$

(28)

Using Pauli-Villas regularization and integral convergence, we obtain:

$$i\prod^{\mu\nu}(q)=(-)\int\frac{d^4p}{(2\pi)^4}\frac{N_{\mu\nu}}{D}$$

(29)

Inside

$$N_{\mu\nu} = tr\left[\gamma_\nu\left(p+q+m\right)\gamma_\nu\left(p+m\right)\right]$$

(30)

$$\frac{1}{D} = \int_0^1 d\alpha \frac{1}{D}$$

(31)

Inside

$$\begin{cases} D = \left[l^2 + \alpha\left(1+\alpha\right)q^2 - m^2 + i\varepsilon\right]^2 \\ l = p + \alpha q \end{cases}$$

(32)

Finally, you can get:

$$\Pi_{\mu\nu}\left(q\right) = -\frac{1}{2\pi^2}\left(q_\mu q_\nu - g_{\mu\nu}q^2\right)\int_0^1 d\alpha\left(1-\alpha\right) \cdot\left\{\ln\left[m^2 - \alpha\left(1-\alpha\right)q^2\right] - \ln\left[m_\alpha^2 - \alpha\left(1-\alpha\right)q^2\right]\right\}$$

(33)

We define:

$$\ln M^2 = \ln m_\alpha^2$$

(34)

Then:

$$\Pi\left(q^2\right) = \frac{1}{2\pi^2}\int_0^1 d\alpha\left(1-\alpha\right)\ln\frac{M^2}{m^2 - \alpha\left(1-\alpha\right)q^2}$$

(35)

Summarizing **Eq. (28)** and **(35):**

$$e_R^2 = e^2\frac{1}{1+\left[\dfrac{e^2}{12\pi^2}\right]\ln\left(\dfrac{M^2}{m^2}\right)} \simeq e^2\left(1-\frac{e^2}{12\pi^2}\ln\frac{M^2}{m^2}\right)$$

(36)

While $M^2$ represents the parameter of the ignorance bound, it confirms that the magnitude of the fluctuation mass of the charge is related to the imaginary mass $M$ .

$$\Delta e^2 = \frac{e^2}{12\pi^2}\ln\frac{M^2}{m^2}$$

(37)

## 2.6 Principles of tropical geometry

Tropical geometry is a variant of algebraic geometry that operates within the framework of "Tropical Mathematics". In tropical geometry, the basic research object is the tropical semiring $\left(\mathbb{R}\cup\{\infty\},\oplus,\odot\right)$ [27], and traditional operations such as addition and multiplication are replaced by "tropical addition" and "tropical multiplication":

$$x\oplus y=\min(x,y),\ x\odot y=x+y \tag{38}$$

The curve defined by a polynomial equation is referred to as an algebraic curve. In traditional algebraic geometry, algebraic curves are described by solving equations; In tropical geometry, these curves are transformed into objects related to 'tropical addition' and 'tropical multiplication,' known as planar tropical curves. For polynomials with two variables:

$$p(x,y)=\bigoplus_{(i,j)} c_{ij}\odot x^i\odot y^j \tag{39}$$

The corresponding planar tropical curve $V(p)$ is the set of all points $(x,y)$ satisfying this transformation. If $p$ is a quadratic polynomial: $p(x,y)=1\odot x^2\oplus 1\odot y^2\oplus 2\odot xy$ the corresponding tropical curve $V(p)$ forms a triangle with vertices at $(0,0),(0,2),(2,0)$:

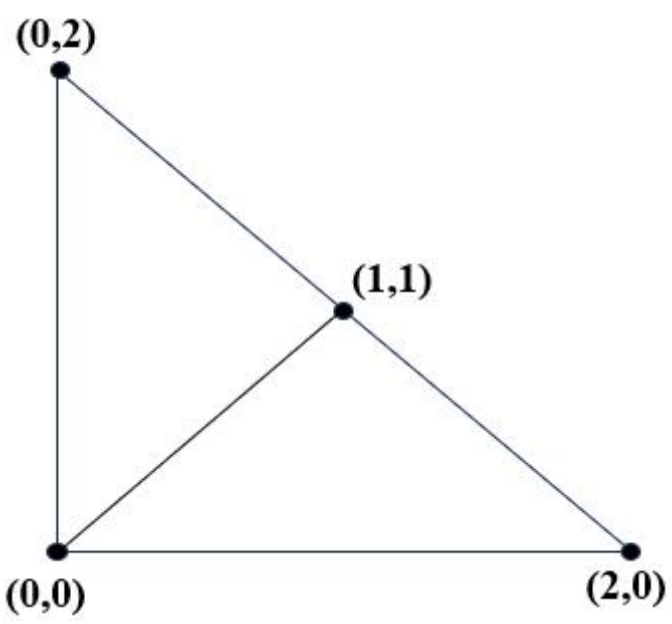

**Fig. 6** Tropical geometry $V\left(1\odot x^2\oplus 1\odot y^2\oplus 2\odot xy\right)$

A tropical matrix is a matrix whose elements are defined using tropical addition and multiplication operations. Unlike conventional matrix operations in traditional linear algebra, tropical matrices follow the rules of tropical mathematics. When discussing eigenvalues of tropical matrices, the concept is analogous to eigenvalues in classical linear algebra but operates under tropical rules. Specifically, each quadratic polynomial can be associated with a $(2\times 2)$ tropical matrix. For example, a polynomial $f(x,y)=2xy$ corresponds to a tropical matrix $A=\begin{bmatrix} 0 & 1 \\ 1 & 0 \end{bmatrix}$, see the Supporting Materials. Similarly, for other forms of polynomials, such as $f(x,y)=x^2+y^2+2xy$ or $f(x,y)=x^2+y^2$, they can also be associated with tropical matrices(Matrix of Quadratic Form) $A=\begin{bmatrix} 1 & 1 \\ 1 & 1 \end{bmatrix}$ and $A=\begin{bmatrix} 1 & 0 \\ 0 & 1 \end{bmatrix}$, respectively. Under this framework, tropical gradient maps can be constructed, showing the relationships and transformation rules between different polynomials and their corresponding tropical matrices:

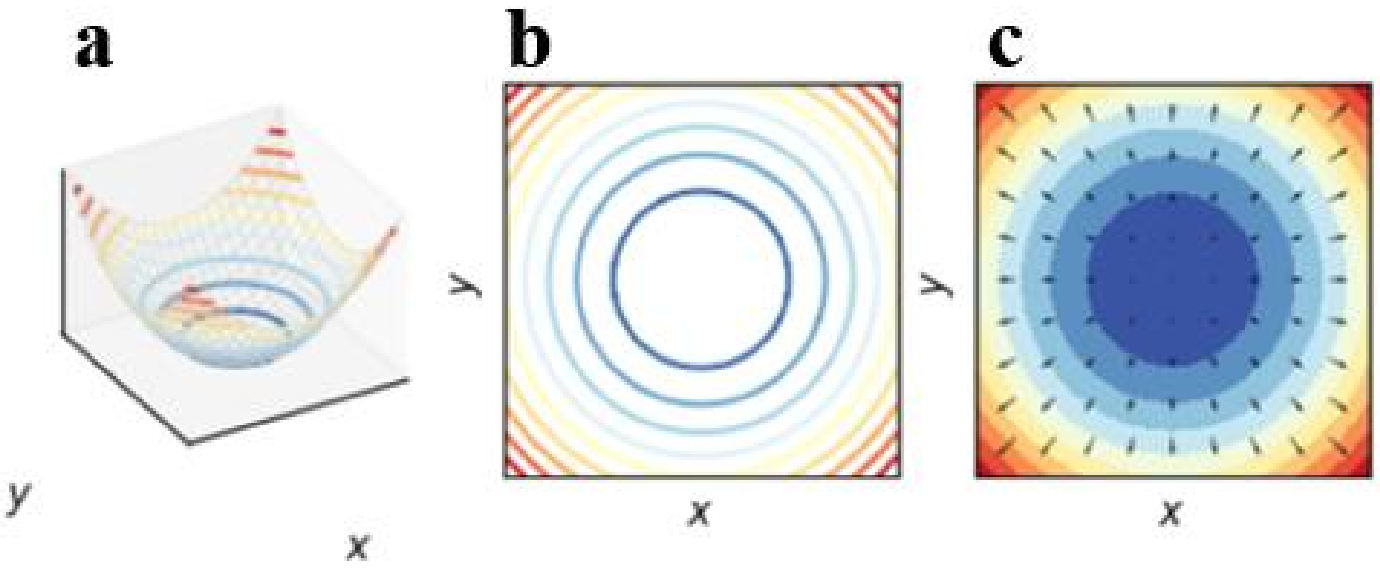

**Fig. 7** Tropical geometry diagram

In **Fig. 7**, **a** presents the 3D surface plot of the quadratic polynomial, **b** shows the 2D contour plot of its function values, and **c** overlays the gradient field vectors on the 2D contour plot. This graph not only provides an intuitive visualization tool for understanding the structure of tropical matrices but also reveals the profound connection between polynomials and matrices in tropical algebra. Tropical geometry represents the equivalent mass in Schrödinger's equation. **(see supplemental material)**

In quantum mechanics, reduced mass is a parameter that describes the "effective" inertia exhibited by particles in certain systems[28]. If we introduce the reduced mass effect in Schrödinger's equation, the reduced mass typically affects the energy spectrum of particles, especially in strong electric fields or non-uniform media. The definition of reduced mass $\bar{M}$ is closely related to the relationship between potential energy and kinetic energy terms in the wave equation. The Hamiltonian operator is:

$$\hat{H}=-\frac{\hbar^2}{2m}\nabla^2+U(\vec{r})$$

(40)

The kinetic energy operator is $\hat{E}_{\mathrm{k}} = -\frac{\hbar^2}{2m}\nabla^2$, where $\nabla^2 = \frac{\partial^2}{\partial x^2} + \frac{\partial^2}{\partial y^2} + \frac{\partial^2}{\partial z^2}$ is the Laplace operator. For the kinetic energy operator, multiply both sides by $\frac{1}{2}x^2$ :

$$\frac{1}{2}\hat{E}_{\mathrm{k}} x^2 = -\frac{\hbar^2}{2m}\nabla^2\left(\frac{1}{2}x^2\right) = -\frac{\hbar^2}{2m}$$

(41)

therefore:

$$\frac{1}{2}\hat{E}_{\mathrm{k}} x_1^2 = -\frac{\hbar^2}{2m_1},\ \frac{1}{2}\hat{E}_{\mathrm{k}} x_2^2 = -\frac{\hbar^2}{2m_2}$$

(42)

let $x_1 = x$, $x_2 = y$, so we have:

$$\frac{1}{2}\left(\hat{E}_{\mathrm{k}} x^2 + \hat{E}_{\mathrm{k}} y^2\right) = -\frac{\hbar^2}{2m_x} - \frac{\hbar^2}{2m_y} = -\frac{\hbar^2}{2}\left(\frac{1}{m_x} + \frac{1}{m_y}\right) = -\frac{\hbar^2}{2}\left(\frac{m_x + m_y}{m_x m_y}\right)$$

(43)

further,

$$\hat{E}_{\mathrm{k}} x^2 + \hat{E}_{\mathrm{k}} y^2 = -\hbar^2\left(\frac{m_x + m_y}{m_x m_y}\right) = -\hbar^2\overline{M}$$

(44)

among them, $\overline{M}$ is the equivalent mass, $\hat{E}_{\mathrm{k}}$ is the kinetic energy operator, and $\hbar$ is the Planck constant. In this form, the reduced mass $\overline{M}$ can be seen as a transformation result generated according to the rules of tropical geometry, representing how the effective mass of particles varies with the state of the system under different conditions. **Eq.43** satisfies the Lorentz transformation, see the Supporting Material.

Therefore, $-\hbar^2\overline{M}$ can be expressed in tropical geometry using the following formula:

$$f(x, y) = \hat{E}_{\mathrm{k}} x^2 + \hat{E}_{\mathrm{k}} y^2$$

(45)

the corresponding tropical matrix is:

$$A=\begin{bmatrix} E_k & 0 \\ 0 & E_k \end{bmatrix}$$

(46)

By integrating tropical geometry with the equivalent mass equation in Schrödinger's equation, we propose a novel approach to re-characterizing the equivalent mass using tropical operations. Representing the transformed mass and potential energy terms as tropical polynomials allows us to not only capture nonlinear effects in the system but also introduce a new mathematical framework for analyzing complex quantum systems.

Future research can further explore the application of tropical geometry in other physical models, particularly in addressing non-uniform and nonlinear quantum systems.

## 3. Results and discussion

### 3.1 Geometric structures

The two-dimensional bilayer $MX_2$ exhibits a typical layered structure, where M atoms and X atoms form a hexagonal arrangement through covalent bonds, while the two layers interact via weak van der Waals forces. Geometric optimization was performed on the initial structures of the two-dimensional bilayer $WS_2$, $WSe_2$, $WTe_2$, and heterojunctions $MoS_2/WSe_2$ and $WSe_2/MoTe_2$. The optimized results are shown in **Fig. 8**. The calculated interlayer distances of $MoS_2/WSe_2$、$WSe_2/MoTe_2$、$WS_2$、$WSe_2$ and $WTe_2$ were 3.837 Å, 3.895 Å, 3.672 Å, 3.740 Å, and 3.900 Å, respectively. Changes in bond length and bond angle reflect variations in the strength and geometric configuration of chemical bonds. Generally, a shorter bond length corresponds to a stronger chemical bond. From the calculation results, it is evident that as the radius of the X atom increases, the M-X bond length also increases. This trend is attributed to the increase in atomic size, which leads to a greater interatomic distance.**(see supplemental material)**

Although changes in bond angles are relatively small, they exhibit identifiable patterns related to the crystal structure's symmetry and atomic interactions. Additionally, phonon spectra of $MoS_2/WSe_2$, $WSe_2/MoTe_2$, $WS_2$, $WSe_2$, and $WTe_2$ reveal no negative frequencies, indicating that these materials have stable structures.

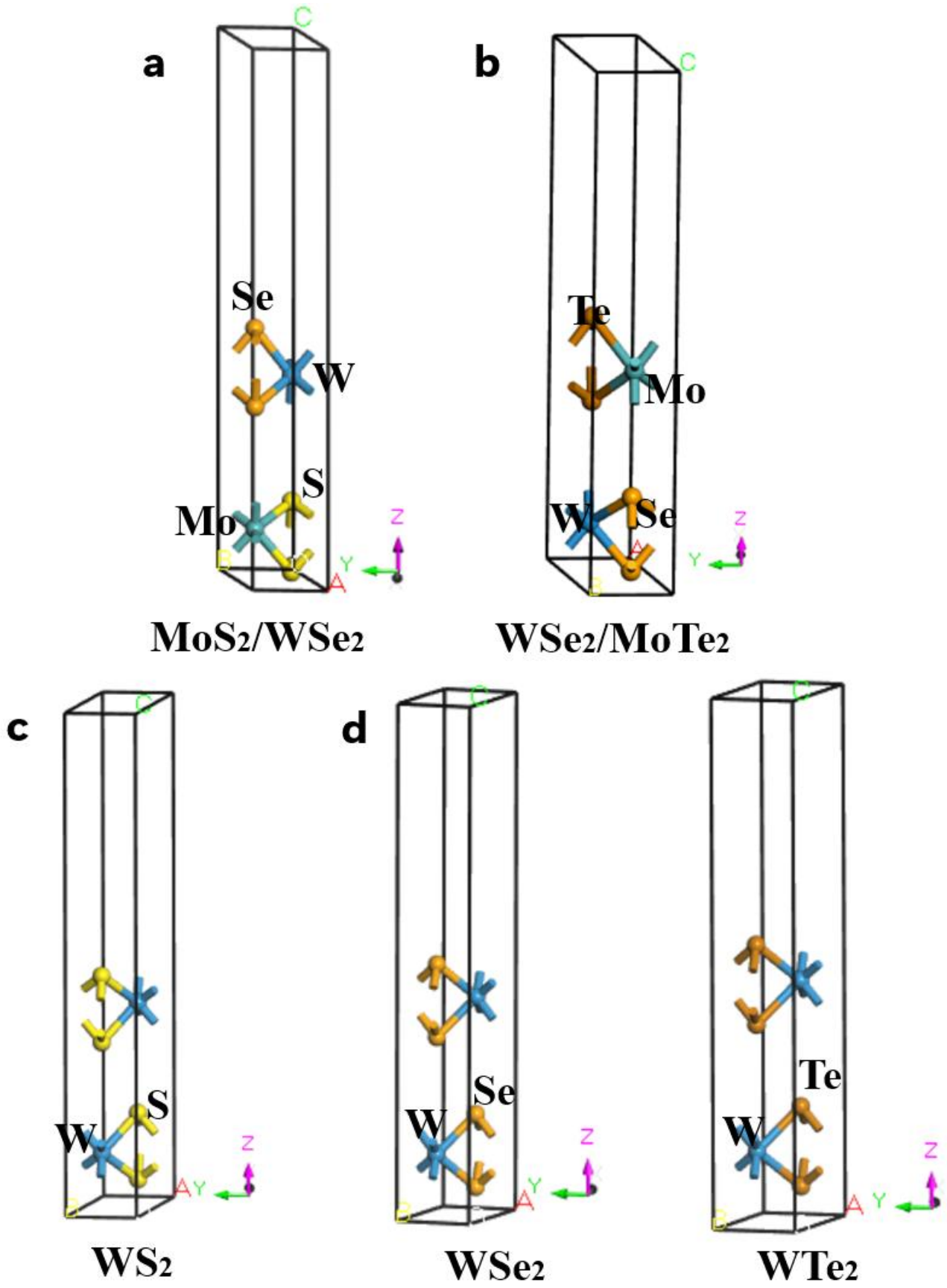

**Fig. 8** Geometrical structures of (a) $MoS_2/WSe_2$, (b) $WSe_2/MoTe_2$, (c) $WS_2$, (d) $WSe_2$ and (e) $WTe_2$ the lattice structure parameter sets are listed in **Table 2**.

**Table 2** Lattice and structural parameters of two-dimensional bilayer $MX_2$

| | Interlayer distance（Å） | Lattice parameters（° / Å） | | | | | | Bond length（Å） | Bond angle（Å） |
|---|---|---|---|---|---|---|---|---|---|
| | | α | β | γ | a | b | c | | |
| $MoS_2$/<br>$WSe_2$ | 3.837 | 90 | 90 | 120 | 3.160 | 3.160 | 24.000 | (Mo-S)<br>2.408<br>(W-Se)<br>2.522 | (S-Mo-S)<br>81.425<br>(Se-W-Se)<br>87.389 |
| $WSe_2$/<br>$MoTe_2$ | 3.895 | 90 | 90 | 60 | 3.282 | 3.282 | 24.000 | (W-Se)<br>2.542<br>(Mo-Te)<br>2.704 | (Se-W-Se)<br>83.587<br>(Te-Mo-Te)<br>91.119 |
| $WS_2$ | 3.672 | 90 | 90 | 120 | 3.153 | 3.153 | 24.000 | 2.413 | 82.069 |
| $WSe_2$ | 3.740 | 90 | 90 | 120 | 3.282 | 3.282 | 24.000 | 2.542 | 83.588 |
| $WTe_2$ | 3.900 | 90 | 90 | 120 | 3.600 | 3.600 | 24.000 | 2.752 | 81.963 |

### 3.2 Band structure, local density of states (LDOS), deformation charge density, and electronic properties

The band structure of the two-dimensional bilayer $MX_2$, calculated using **DFT**, is shown in **Fig. 9**. The results indicate that all compounds exhibit semiconductor properties, with bandgap widths varying within a specific range. Molecular dynamics simulation shows that the two-dimensional bilayer $MX_2$ is stable. The results are shown in the supplemental material. The bandgap width of $MoS_2/WSe_2$ is 0.934 eV, $WSe_2/MoTe_2$ is 0.334 eV, $WS_2$ is 1.838 eV, $WSe_2$ is 1.496 eV, and $WTe_2$ is 0.917 eV, respectively. From the band structure, the energy bands near the top of the valence band and the bottom of the conduction band exhibit distinct dispersion relations. These variations reflect differences in the effective mass and motion characteristics of electrons in different directions.

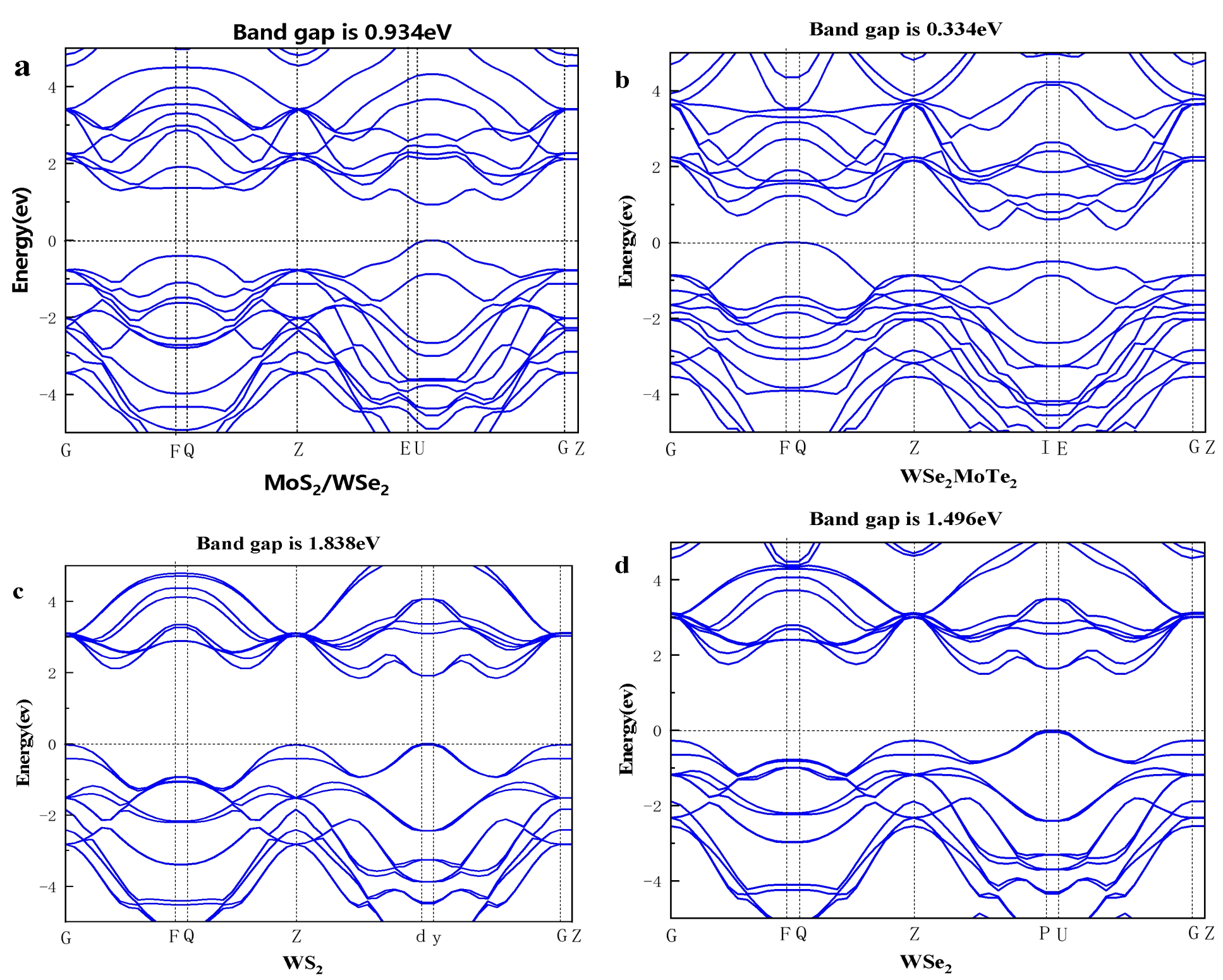

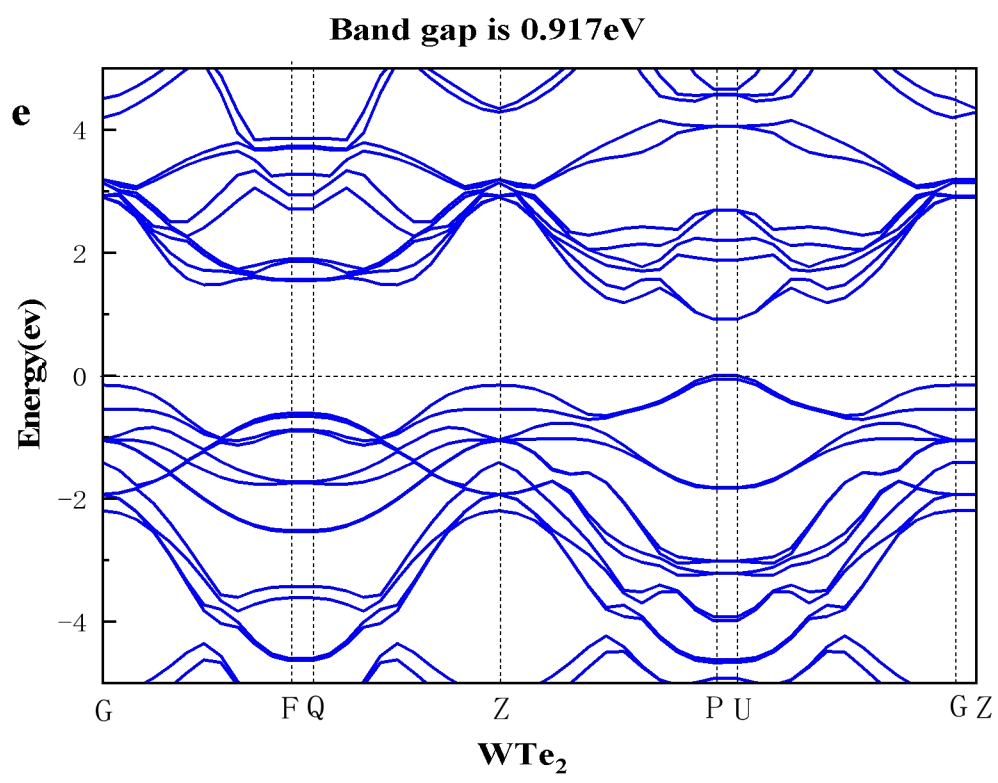

**Fig. 9** Band structure of (a) $MoS_2/WSe_2$, (b) $WSe_2/MoTe_2$, (c) $WS_2$, (d) $WSe_2$ and (e)$WTe_2$.

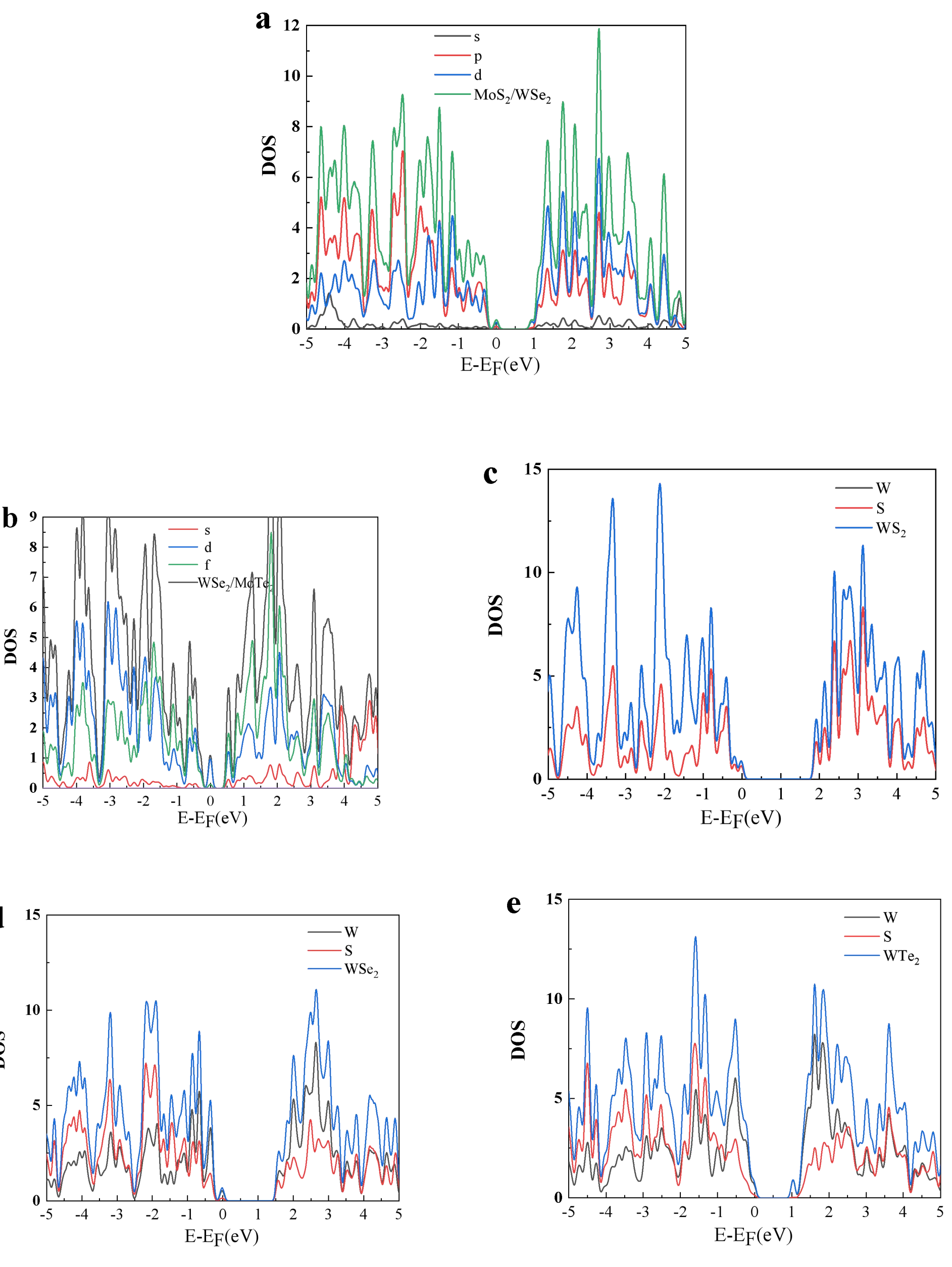

**Fig. 10** density of states (DOS) of (a) $MoS_2/WSe_2$, (b) $WSe_2/MoTe_2$, (c) $WS_2$, (d)

$WSe_2$ and (e) $WTe_2$

**Fig. 10** shows the total DOS and PDOS of the two-dimensional bilayer $MX_2$. The total DOS reveals that the primary peaks of the electron density distribution in the conduction bands of the $MoS_2/WSe_2$ and $WSe_2/MoTe_2$ heterojunctions are located in the negative energy region, suggesting that these structures are relatively stable. Near the Fermi level, the DOS is primarily contributed by the d orbitals of M and the p orbitals of X. In Mo-based compounds, the d orbitals of Mo contribute significantly in the lower energy region, whereas the p orbitals of X dominate in the higher energy region. A similar orbital contribution pattern is observed in W-based compounds; however, due to the differences in the atomic structure and electronic properties of W compared to Mo, the specific distribution of DOS varies accordingly. The DOS further provides insight into the contributions of different atomic orbitals to electronic states and their mutual hybridization. For instance, in $MoS_2$, the $d_{x^2-y^2}$ and $d_z$ orbitals of Mo exhibit strong hybridization with the $p_x$ and $p_\gamma$ orbitals of S within a specific energy range. This orbital hybridization plays a crucial role in chemical bond formation and the stability of electronic structures. Additionally, the energy levels of bilayer $WSe_2$, $WS_2$, and $WTe_2$ intersect with one or more energy bands, indicating high electron mobility. Therefore, these materials hold significant potential as candidates for electronic device applications.

To analyze the electronic properties of the bonds, we used the deformation charge density (**Fig. 11**) of the Mo-S chemical bonds in $MoS_2/WSe_2$ (**Fig. 11a**)、 the W-Se chemical bonds in $WSe_2/MoTe_2$(**Fig. 11b**)、the W-S chemical bonds in $WS_2$ **Fig. 11c**)、 the W-Se chemical bonds in $WSe_2$(**Fig. 11d**) and the W-Te chemical bonds in $WTe_2$ (**Fig. 11e**). The electron distributions are represented by a color scale, where blue and red areas indicate an increase and decrease in electron density, respectively. The Te, S, and Se atoms primarily appear in blue, signifying a positive deformation charge density and a significant charge accumulation in these regions of the structure. Conversely, the W and Mo atoms predominantly appear in red, indicating a negative deformation charge density and a substantial charge divergence.

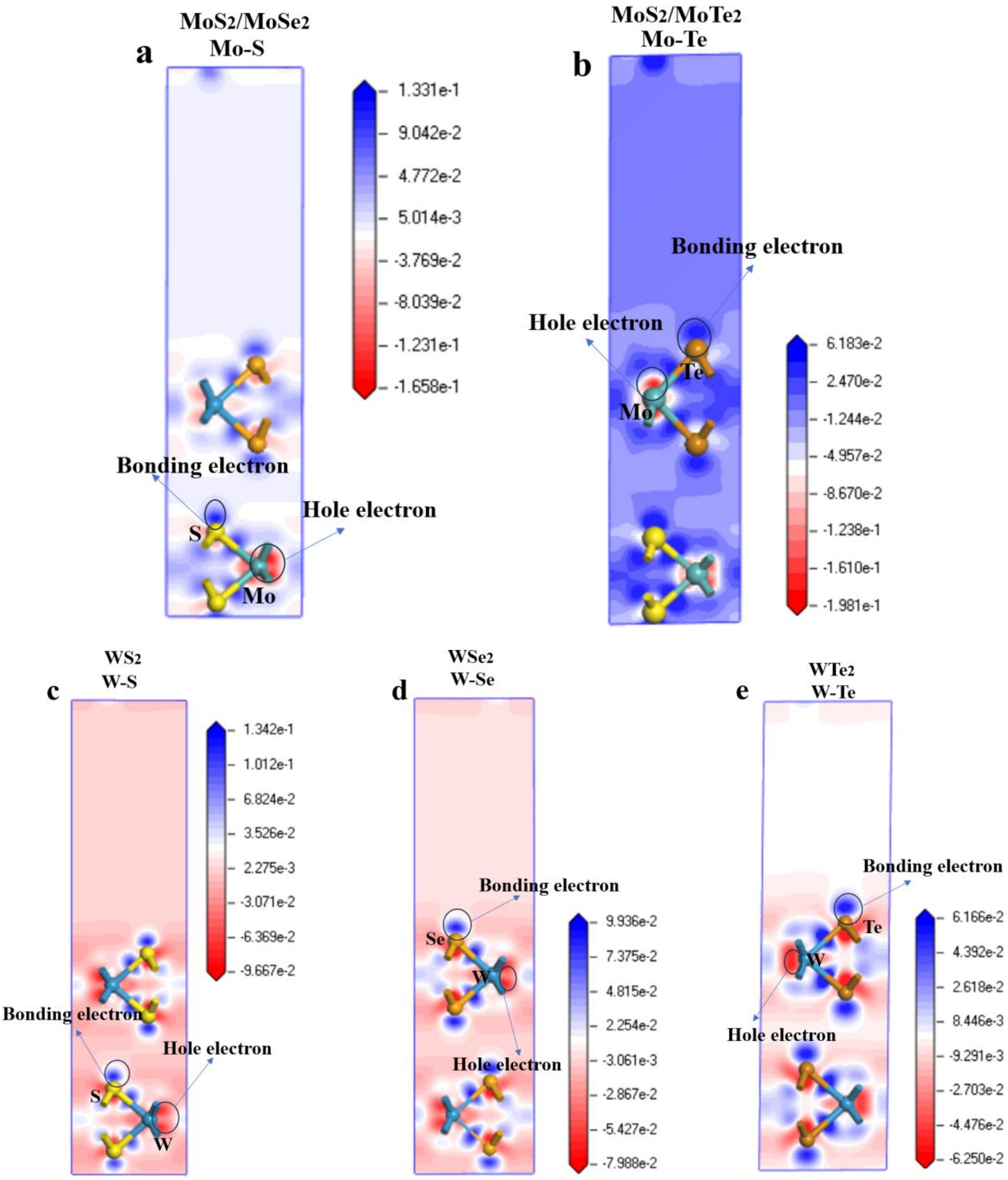

**Fig. 11** Deformation charge densities of (a) $MoS_2/WSe_2$, (b) $WSe_2/MoTe_2$, (c)$WS_2$, (d) $WSe_2$ and (e) $WTe_2$

The established BBC model was used to convert the Hamiltonian values into bonding values (**Table 3**).

**Table 3** Deformation charge density $\delta\rho\left(\overrightarrow{r_{ij}}\right)$ and deformation bond energy $\delta V_{bc}\left(\overrightarrow{r_{ij}}\right)$, as obtained from the BBC model.

$$\left(\varepsilon_0 = 8.85\times10^{-12}C^2N^{-1}m^{-2}, e = 1.60\times10^{-19}C, \left|\vec{r}_{ij}\right| \approx d_{ij}/2\right)$$

| | $MoS_2/$ | $WSe_2/$ | $WS_2$ | $WSe_2$ | $WTe_2$ |
|---|---|---|---|---|---|

| | $WSe_2$ | $MoTe_2$ | | | |
|---|---|---|---|---|---|
| $r_{ij}$ ( Å ) | 1.20<br>Mo-S | 1.27<br>W-Se | 1.21<br>W-S | 1.27<br>W-Se | 1.38<br>W-Te |
| $r_i$ ( Å ) | 1.05<br>(S) | 1.20<br>(Se) | 1.05<br>(S) | 1.20<br>(Se) | 1.38<br>(Te) |
| $\vec{r}_j$ ( Å ) | 1.54<br>(Mo) | 1.62<br>(W) | 1.62<br>(W) | 1.62<br>(W) | 1.62<br>(W) |
| $\delta\rho^{Bonding-electron}\left(\vec{r}_i\right)\left(e/\mathring{A}^3\right)$ | 0.1331<br>(S) | 0.1038<br>(Se) | 0.1342<br>(S) | 0.0994<br>(Se) | 0.0676<br>(Te) |
| $\delta\rho^{Hole-electron}\left(\vec{r}_j\right)\left(e/\mathring{A}^3\right)$ | -0.1658<br>(Mo) | -0.0822<br>(W) | -0.0967<br>(W) | -0.0799<br>(W) | -0.0600<br>(W) |
| $\delta V_{bc}^{bonding}\left(\vec{r}_{ij}\right)(eV)$ | -0.5590 | -0.3549 | -0.3795 | -0.3303 | -0.2361 |
| $\delta V_{bc}^{Antibonding}\left(\vec{r}_i\right)(eV)$ | 0.1626<br>(S) | 0.1928<br>(Se) | 0.1653<br>(S) | 0.1768<br>(Se) | 0.1644<br>(Te) |
| $\delta V_{bc}^{Antibonding}\left(\vec{r}_j\right)(eV)$ | 1.7120<br>(Mo) | 0.5421<br>(W) | 0.7502<br>(W) | 0.5122<br>(W) | 0.2888<br>(W) |

The deformation charge density and electronic radius were used to calculate the deformation bond energies from **Eq. 7** as follows: -0.5590, -0.3549, -0.3795, -0.3303 and -0.2361 eV for Mo-S in $MoS_2/WSe_2$, W-Se in $WSe_2/MoTe_2$, W-S in $WS_2$, W-Se in $WSe_2$ and W-Te in $WTe_2$, respectively.

### 3.3 Non-Hermitian bonding of $MoS_2/MoSe_2$、$MoS_2/MoTe_2$、$WSe_2$/Mo-Te、$WS_2$、$WSe_2$ and $WTe_2$

$$M(z)=\frac{az+b}{cz+d}\rightarrow[M]=\begin{bmatrix} a & b \\ c & d \end{bmatrix} \tag{48}$$

The Möbius transform is a fractional linear function that maps point $z$ on a complex plane to a unique point, where $a$, $b$, $c$, and $d$ are plurals, and $ad-bc\neq 0$.To use **Eq. 11** for a point $W = (X, Y, Z)$ in real space, we first normalize the coordinates $R_1(x_1,y_1,z_1)$ and $R_2(x_2,y_2,z_2)$ of two atoms in real space. The distance $l$ between them is:

$$\sqrt{(x_1-x_2)^2+(y_1-y_2)^2+(z_1-z_2)^2}=l \tag{49}$$

Therefore:

$$\sqrt{\frac{(x_1-x_2)^2}{l^2}+\frac{(y_1-y_2)^2}{l^2}+\frac{(z_1-z_2)^2}{l^2}}=1 \tag{50}$$

Order:

$$\begin{cases} X=\dfrac{x_1-x_2}{l} \\ Y=\dfrac{y_1-y_2}{l} \\ Z=\dfrac{z_1-z_2}{l} \end{cases} \tag{51}$$

Point $W = (X, Y, Z)$ in real space satisfies $\sqrt{X^2+Y^2+Z^2}=1$.

For a point $w$ in the complex number set $C$, $w = x + iy$, $\mathrm{Re}(w)=x$, and $\mathrm{Im}(w)=y$. As the projection point of $W$, $w$ satisfies $x=\dfrac{X}{1-Z}$ or $y=\dfrac{Y}{1-Z}$. The projections of the atomic coordinates from real space onto a two-dimensional complex plane are listed in **Table 4**.

**Table 4** Spatial coordinates of chemical bonds between $MoS_2/WSe_2$, $WSe_2/MoTe_2$, $WS_2$, $WSe_2$ and $WTe_2$ and their projection points on the complex plane.

| | Atom 1 (coordinates) | Atom 2 (coordinates) | X | Y | Z | W |
|---|---|---|---|---|---|---|
| $MoS_2/WSe_2$ | Mo (0.91,1.58,2.08) | S (1.82,0,3.65) | 0.1430 | 0.4312 | 0.4258 | 0.2491+0.7509$i$ |
| $WSe_2/MoTe_2$ | W (1.91,3.30,2.22) | Se (0.96,1.65,3.91) | 0.1393 | 0.4201 | 0.4407 | 0.2490+0.7510$i$ |
| $WS_2$ | W | S | 0.1415 | 0.4265 | 0.4320 | 0.2491+0.7509$i$ |
| | (1.82,0,9.58) | (0.91,1.58,11,17) | | | | |
| $WSe_2$ | W<br>(1.90,0,10.04) | Se<br>(0.95,1.63,11.74) | 0.1399 | 0.4120 | 0.4481 | 0.2536+0.7464*i* |
| $WTe_2$ | W<br>(2.08,0,10.85) | Te<br>(1.04,1.80,12.65) | 0.1430 | 0.4268 | 0.4285 | 0.2503+0.7497*i* |

Based on **Eq. 49**, the projection points of Mo-S in $MoS_2/WSe_2$, W-Se in $WSe_2/MoTe_2$, W-S in $WS_2$, W-Se in $WSe_2$ and W-Te in $WTe_2$ in a two-dimensional complex plane are 0.2491 + 0.7509*i*, 0.2490 + 0.7510*i*, 0.2491 + 0.7509*i*, 0.2436 + 0.7464*i* and 0.2503 + 0.7497*i*, respectively.

The elements of the matrix $\begin{bmatrix} a & b \\ c & d \end{bmatrix}$ in the Möbius transformation are complex numbers. To determine the projection of chemical bond energy, according to **Table 4**, we define the bond energy coefficients $V_{11}$, $V_{22}$, and $V_{12}$ as:

$$\begin{cases} V_{11} = \dfrac{1}{4\pi\varepsilon_0}\dfrac{e^2}{2\left|\vec{r}_i\right|}\displaystyle\int d^3r\int d^3r\,\delta\rho^{Bonding-electron}\left(\vec{r}_i\right)\delta\rho^{Bonding-electron}\left(\vec{r}_i\right) = \delta V_{bc}^{Antibonding}\left(\vec{r}_i\right) \\ V_{22} = \dfrac{1}{4\pi\varepsilon_0}\dfrac{e^2}{2\left|\vec{r}_j\right|}\displaystyle\int d^3r\int d^3r\,\delta\rho^{Hole-electron}\left(\vec{r}_j\right)\delta\rho^{Hole-electron}\left(\vec{r}_j\right) = \delta V_{bc}^{Antibonding}\left(\vec{r}_j\right) \\ V_{12} = \dfrac{1}{4\pi\varepsilon_0}\dfrac{e^2}{2\left|\vec{r}_i-\vec{r}_j\right|}\displaystyle\int d^3r\int d^3r\,\delta\rho^{Bonding-electron}\left(\vec{r}_{ij}\right)\delta\rho^{Hole-electron}\left(\vec{r}_{ij}\right) = \delta V_{bc}^{bonding}\left(\vec{r}_{ij}\right) \end{cases}$$

(52)

Substituting **Eq. 49** and *Z*=*w*=*x*+*iy* into the matrix elements of the Möbius transformation yields:

$$\begin{bmatrix} a & b \\ c & d \end{bmatrix} = \begin{bmatrix} V_{12}i & V_{11}i \\ V_{21}i & V_{22}i \end{bmatrix}$$

(53)

Substituting **Eq. 50** into **Eq. 48**, we obtain:

$$b = V_{11} = \delta V_{bc}^{Antibonding}\left(\vec{r}_i\right), d = V_{22} = \delta V_{bc}^{Antibonding}\left(\vec{r}_j\right), a = c = V_{12} = V_{21} = \delta V_{bc}^{bonding}\left(\vec{r}_{ij}\right)$$

(54)

Substituting **Eq. 51** into **Eq. 48** yields the Tan and Bo transformation[29] with the chemical bond energy as a coefficient:

$$F(z) = \frac{V_{12}iZ + V_{11}i}{V_{21}iZ + V_{22}i}$$

(55)

The chemical bond energy is used as a coefficient to determine the eigenvalues of the Hamiltonian energy, which are then transformed into energy projections in complex space through the Möbius transformation. The chemical bond energy coefficients are calculated by substituting Eq. 49 into Eq. 52, and the result is further substituted into Eq. 55 to obtain the Möbius transformation of $MoS_2/WSe_2$, $WSe_2/MoTe_2$, $WS_2$, $WSe_2$ and $WTe_2$.

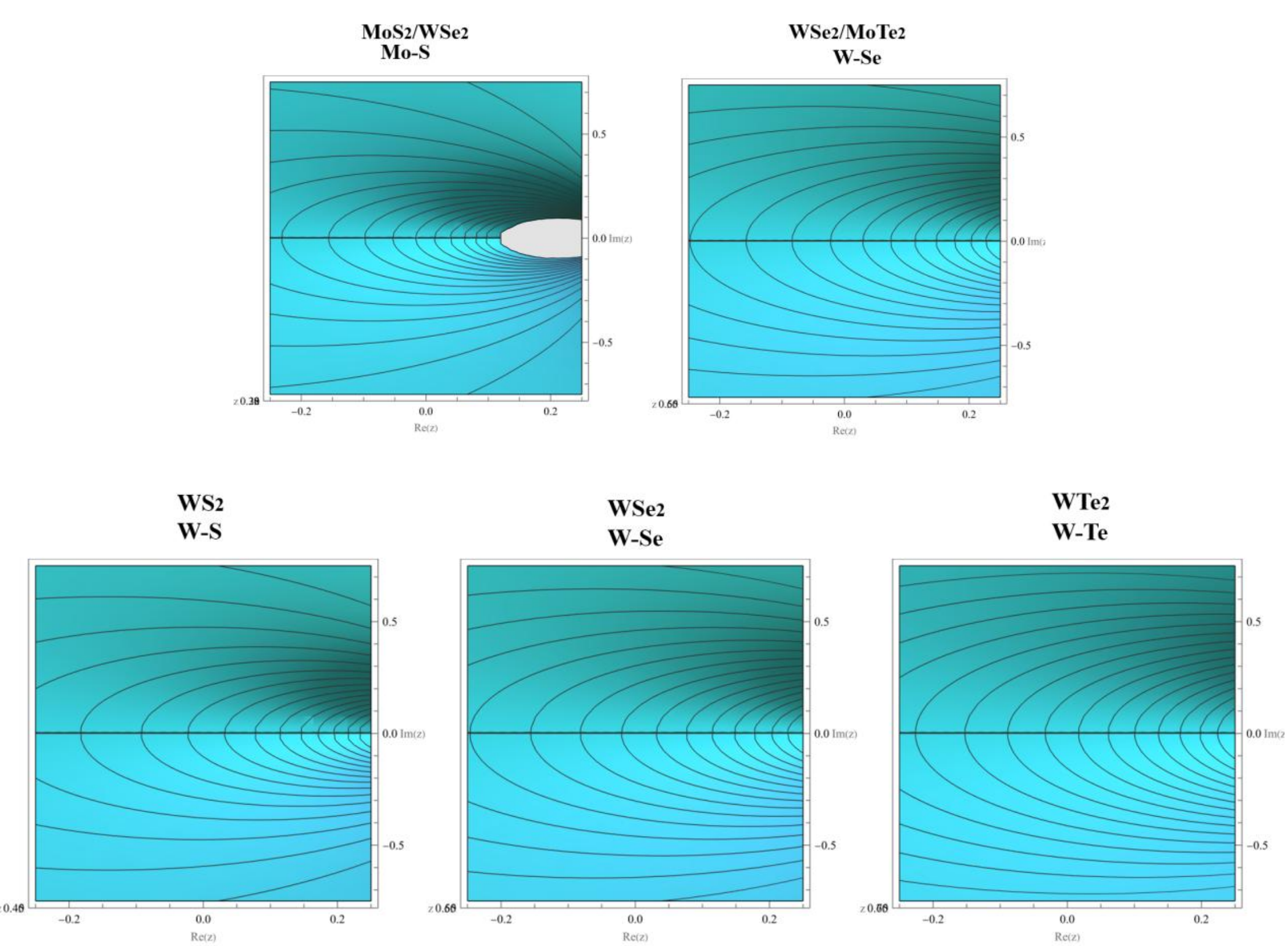

**Fig. 12** Non-Hermitian chemical bond of $MoS_2/WSe_2$, $WSe_2/MoTe_2$, $WS_2$, $WSe_2$ and $WTe_2$

The calculations using the $Z=w=x+iy$ function, illustrated in **Fig.12**, are as follows:

$$F(z)=\frac{-0.5590i\cdot w+0.1626i}{-0.5590i\cdot w+1.7120i}$$ , $w$ =0.25+0.75$i$ of Mo-S in $MoS_2/WSe_2$;

$$F(z)=\frac{-0.3549i\cdot w+0.1928i}{-0.3549i\cdot w+0.5421i}$$ , $w$ =0.25+0.75$i$ of W-Se in $WSe_2/MoTe_2$;

$$F(z)=\frac{-0.3795i\cdot w+0.1653i}{-0.3795i\cdot w+0.7502i}$$ , $w$ =0.25+0.75$i$ of W-S in $WS_2$;

$$F(z)=\frac{-0.3303i\cdot w+0.1768i}{-0.3303i\cdot w+0.5122i}$$ , $w$ =0.25+0.75$i$ of W-Se in $WSe_2$;

$$F(z)=\frac{-0.2361i\cdot w+0.1644i}{-0.2361i\cdot w+0.2888i}$$ , $w$ =0.25+0.75$i$ of W-Te in $WTe_2$.

### 3.4 Conformal transformation

Each $R(x+yi)$ can be represented in the following form:

$$R(x+yi)=P(x+yi)/Q(x+yi)$$

(56)

and

$$R(x+yi)=X+Yi,$$

$$R(x-yi)=X-Yi$$

$$(A+Bi)/(C+Di)=(A+Bi)(C-Di)/(C+Di)(C-Di)=\frac{AC+BD}{C^2+D^2}+\frac{BC-AD}{C^2+D^2}i$$

(57)

So $R(x+yi)$ can be expressed as:

$$R(x+yi)=\frac{A+Bi}{C+Di}=\frac{AC+BD}{C^2+D^2}+\frac{BC-AD}{C^2+D^2}i$$

(58)

Similarly, $R(x-yi)$ can be expressed as:

$$R(x-yi)=\frac{A-Bi}{C-Di}=\frac{AC+BD}{C^2+D^2}-\frac{BC-AD}{C^2+D^2}i$$

(59)

Therefore, each transformation can be expressed as a complex number.
If A, B, C, and D are real numbers and measurable physical quantities, their relationship follows de Moivre's theorem:

$$\begin{aligned}&r_1(\cos\theta_1+i\sin\theta_1)\times r_2(\cos\theta_2+i\sin\theta_2)\times\cdots\times r_n(\cos\theta_n+i\sin\theta_n)\\&=r_1r_2\cdots r_n\{\cos(\theta_1+\theta_2+\cdots+\theta_n)+\sin(\theta_1+\theta_2+\cdots+\theta_n)\}\end{aligned}$$

(60)

where $r_1=r_2=\cdots=r_n=1$， $\theta_1=\theta_2=\cdots=\theta_n=\theta$, obtain

$$(\cos\theta+i\sin\theta)^n=\cos n\theta+i\sin n\theta$$

(61)

where $n$ is any positive integer, if $z=r(\cos\theta+i\sin\theta)$, then $1/z=(\cos\theta-i\sin\theta)/r$ .

Another one

$$
\begin{aligned}
\left(\cos\theta + i\sin\theta\right)^{-n} &= \left(\cos\theta - i\sin\theta\right)^{n} \\
&= \left\{\cos\left(-\theta\right) + i\sin\left(-\theta\right)\right\}^{n} \\
&= \cos\left(-n\theta\right) + i\sin\left(-n\theta\right)
\end{aligned}
$$

(62)

So, the de Moivre theorem holds for all positive or negative integer values of $n$.

Now let ABCD be the measurable value of energy, and based on the wave function $e^{i\theta} = \cos\theta + i\sin\theta$ , we can directly convert the energy value into the form of a wave function：

$$
r\left(\cos\theta + i\sin\theta\right) = \begin{cases} \dfrac{A+Bi}{C+Di} = \dfrac{AC+BD}{C^2+D^2} + \dfrac{BC-AD}{C^2+D^2} i \ (B > D) \\ \dfrac{A-Bi}{C-Di} = \dfrac{AC+BD}{C^2+D^2} - \dfrac{BC-AD}{C^2+D^2} i \ (D > B) \end{cases}
$$

$$
r\left(\cos\theta - i\sin\theta\right) = \begin{cases} \dfrac{A+Bi}{C+Di} = \dfrac{AC+BD}{C^2+D^2} + \dfrac{BC-AD}{C^2+D^2} i \ (D > B) \\ \dfrac{A-Bi}{C-Di} = \dfrac{AC+BD}{C^2+D^2} - \dfrac{BC-AD}{C^2+D^2} i \ (B > D) \end{cases}
$$

(63)

Multiplying infinite amounts of energy (phase angle) into waveform for the same amount of energy

$$
\begin{aligned}
&\left(\frac{A+Bi}{C+Di}\right) \times \left(\frac{A+Bi}{C+Di}\right) \times \cdots \times \left(\frac{A+Bi}{C+Di}\right) \\
&= \left(\frac{AC+BD}{C^2+D^2} + \frac{BC-AD}{C^2+D^2} i\right)^{n} \\
&= r^{n}\left(\cos\theta + i\sin\theta\right)^{n} \\
&= r^{n}\left(\cos n\theta + i\sin n\theta\right) \\
&= r^{n} e^{in\theta}
\end{aligned}
$$

(64)

For different energies

$$
\begin{aligned}
&\left(\frac{A_1+B_1 i}{C_1+D_1 i}\right) \times \left(\frac{A_2+B_2 i}{C_2+D_2 i}\right) \times \cdots \times \left(\frac{A_k+B_k i}{C_k+D_k i}\right) \\
&= \left(\frac{A_1C_1+B_1D_1}{C_1^2+D_1^2} + \frac{B_1C_1-A_1D_1}{C_1^2+D_1^2} i\right) \times \cdots \times \left(\frac{A_kC_k+B_kD_k}{C_k^2+D_k^2} + \frac{B_kC_k-A_kD_k}{C_k^2+D_k^2} i\right) \\
&= r_1\left(\cos\theta_1 + i\sin\theta_1\right) \times r_2\left(\cos\theta_2 + i\sin\theta_2\right) \times \cdots \times r_k\left(\cos\theta_k + i\sin\theta_k\right) \\
&= r_1 r_2 \cdots r_k \left\{\cos\left(\theta_1+\theta_2+\cdots+\theta_k\right) + \sin\left(\theta_1+\theta_2+\cdots+\theta_k\right)\right\}
\end{aligned}
$$

(65)

Let ABCD be the measurable value of energy, based on the energy values calculated in **Table 3**, that is $B=\delta V_{bc}^{Antibonding}\left(\vec{r_i}\right)$, $D=\delta V_{bc}^{Antibonding}\left(\vec{r}_j\right)$, $A=C=\delta V_{bc}^{bonding}\left(\vec{r_{ij}}\right)$, for example, in the heterojunction $MoS_2/WSe_2$, $B=0.1626$, $D=1.7120$, $A=C=-0.5590$ be substituted into **Eq. 63** :

$$\frac{A+Bi}{C+Di}=\frac{AC+BD}{C^2+D^2}+\frac{BC-AD}{C^2+D^2}i=0.1822+0.2670i$$

Therefore, the measurable value of energy is represented as a complex number. Now, convert the energy value directly into the form of a wave function $r\left(\cos\theta+i\sin\theta\right)=re^{i\theta}$. The modulus $r$ and radiation angle $\theta$ are calculated using the following formula:

$$r=\sqrt{x^2+y^2}$$

$$\theta\begin{cases} x>0,\arctan(y/x) \\ x=0,y>0; \\ x=0,y<0; \\ x<0,\arctan(y/x)+\pi \end{cases}$$

(66)

that is

$$0.1822+0.2670i=0.3233(\cos 0.97+i\sin 0.97)$$

Substitute into **Eq. 64**, the wave function $r^n e^{in\theta}=0.3233^n\left(\cos 0.97n+i\sin 0.97n\right)$ can be obtained.

Now, substitute the energy values of six different substances, $MoS_2/WSe_2$, $WSe_2/MoTe_2$, $WS_2$, $WSe_2$, and $WTe_2$, into **Eq. 63** obtained:

$$\begin{cases} 0.1822+0.2670i=0.3233(\cos 0.97+i\sin 0.97) \\ 0.5490+0.2953i=0.6233\left(\cos 0.49+i\sin 0.49\right) \\ 0.3792+0.3140i=0.4924\left(\cos 0.69+i\sin 0.69\right) \\ 0.5375+0.2982i=0.6147\left(\cos 0.51+i\sin 0.51\right) \\ 0.7418+0.2111i=0.7713\left(\cos 0.28+i\sin 0.28\right) \end{cases}$$

as shown in **Table 5.**

**Table 5** The conformal transformation of energy obtains the argument $\theta_k$ and amplitude $r_k$ of the wave function

| | B | D | A=C | $\dfrac{AC+BD}{C^2+D^2}+\dfrac{BC-AD}{C^2+D^2}i$ | $\theta$(rad) | $r$ |
|---|---|---|---|---|---|---|
| $MoS_2/WSe_2$ | 0.1626 | 1.7120 | -0.5590 | $0.1822+0.2670i$ | 0.97 | 0.3233 |
| $WSe_2/MoTe_2$ | 0.1928 | 0.5421 | -0.3549 | $0.5490+0.2953i$ | 0.49 | 0.6233 |
| $WS_2$ | 0.1653 | 0.7502 | -0.3795 | $0.3792+0.3140i$ | 0.69 | 0.4924 |
| $WSe_2$ | 0.1768 | 0.5122 | -0.3303 | $0.5375+0.2982i$ | 0.51 | 0.6147 |
| $WTe_2$ | 0.1644 | 0.2888 | -0.2361 | $0.7418+0.2111i$ | 0.28 | 0.7713 |

### 3.5 Fluctuation of electron mass and reduced mass

#### 3.5.1 Fluctuation of electron mass

From the **Eq. 28** and **Eq. 35** summary:

$$e_R^2 = e^2 \frac{1}{1+\left[\dfrac{e^2}{12\pi^2}\right]\ln\left(\dfrac{M^2}{m^2}\right)} \simeq e^2\left(1-\frac{e^2}{12\pi^2}\ln\frac{M^2}{m^2}\right)$$

While $M^2$ representing the parameter of the ignorance bound, which confirms that the magnitude of the fluctuation mass of the charge is related to the imaginary mass $M$.

$$\Delta e^2 = \frac{e^2}{12\pi^2}\ln\frac{M^2}{m^2}$$

Among them, electronic quality $m = 9.10956\times10^{-31}\,kg$

**Fig. 13** shows the atomic positions, types, and corresponding charges of six two-dimensional bilayer structures

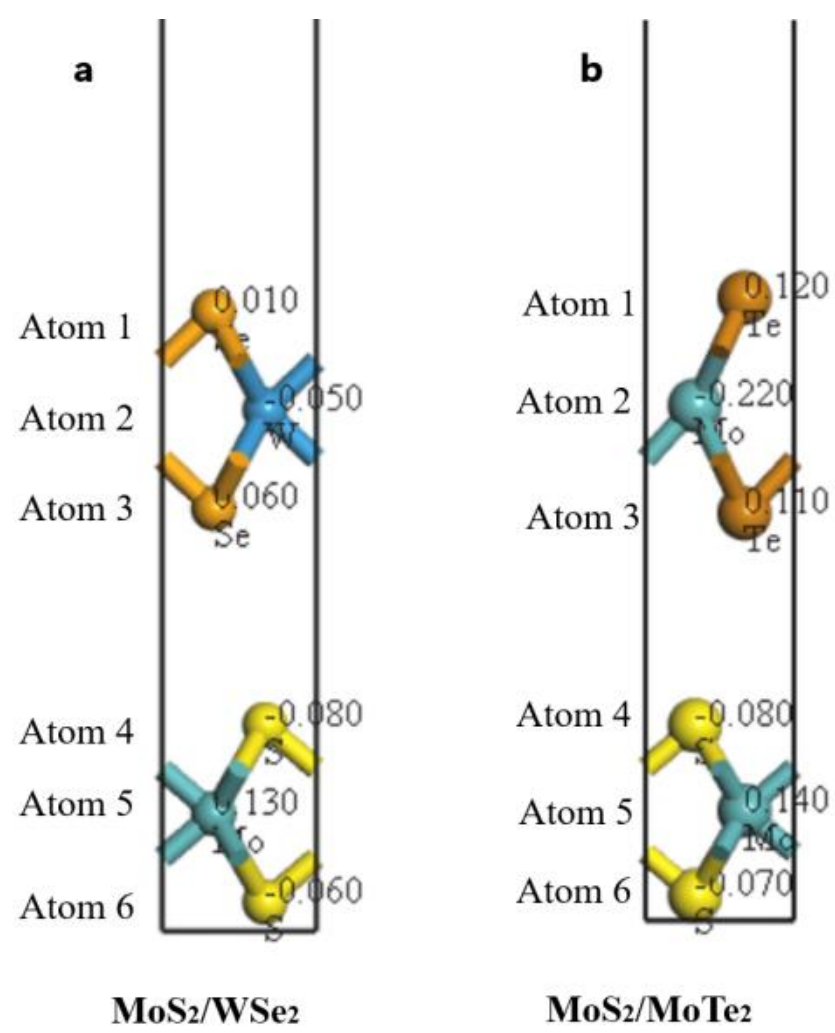

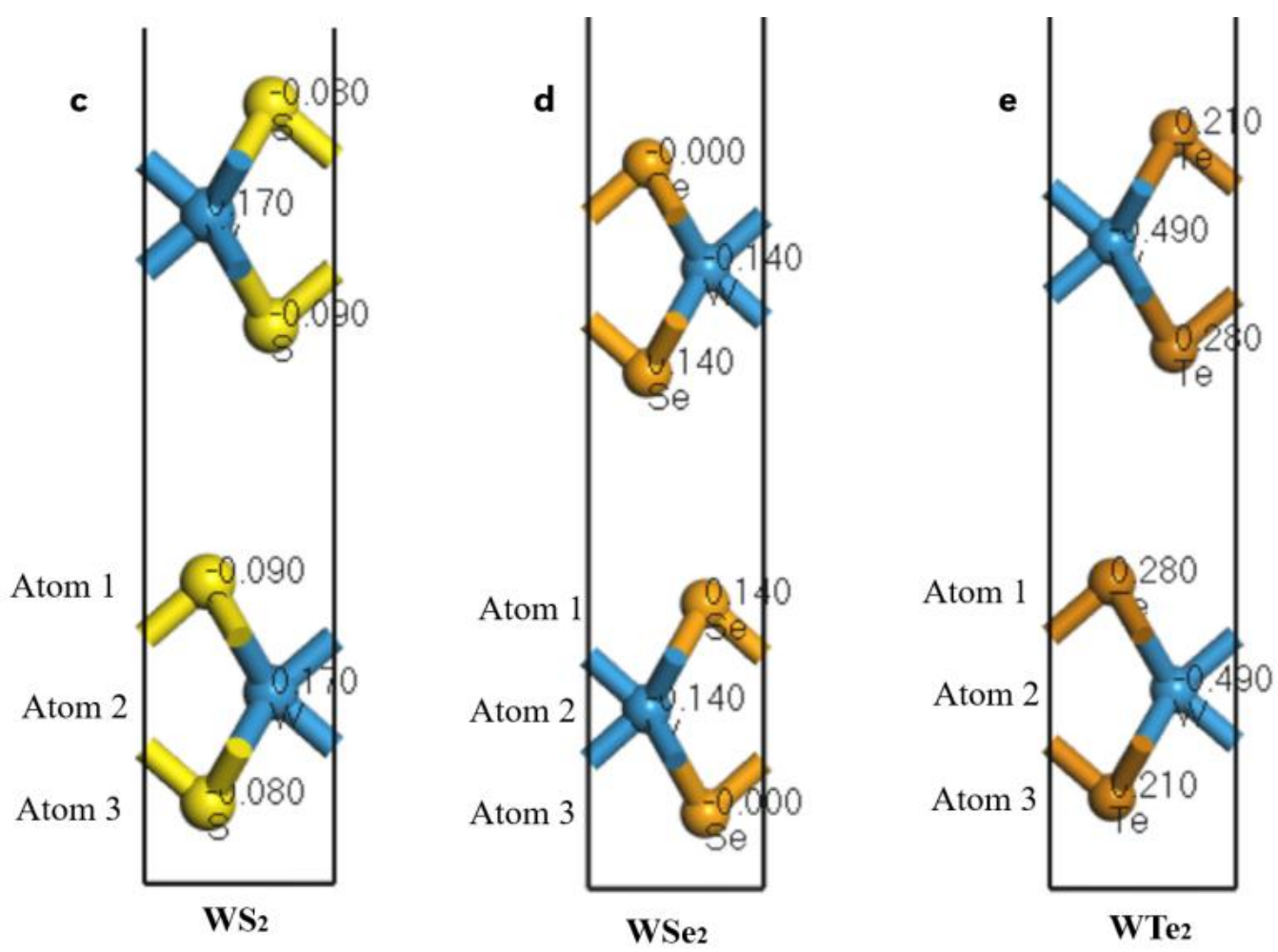

**Fig. 13** Charge transfer of $MoS_2/WSe_2$, $WSe_2/MoTe_2$, $WS_2$, $WSe_2$ and $WTe_2$

Based on **Fig. 13**, substitute the charges of each atom into **Eq. 37** to obtain the mass fluctuations of each atom, as shown in **Table 6**

**Table 6** Fluctuation of electron mass M and charge transfer $\Delta e$

| Structure | | element | $\Delta e$ | $M$ |
|---|---|---|---|---|
| $MoS_2/WSe_2$ | Atom1 | Se | 0.010 | 9.16E-31 |
| | Atom2 | W | -0.050 | 1.06E-30 |
| | Atom3 | Se | 0.060 | 1.13E-30 |
| | Atom4 | S | -0.080 | 1.33E-30 |
| | Atom5 | Mo | 0.130 | 2.48E-30 |
| | Atom6 | S | -0.060 | 1.13E-30 |
| $WSe_2/MoTe_2$ | Atom1 | Te | 0.160 | 4.14E-30 |
| | Atom2 | Mo | -0.380 | 4.67E-27 |
| | Atom3 | Te | 0.180 | 6.19E-30 |
| | Atom4 | Se | 0.170 | 5.03E-30 |
| | Atom5 | W | -0.130 | 2.48E-30 |
| | Atom6 | Se | 0 | 9.11E-31 |
| $WS_2$ | Atom1 | S | -0.090 | 1.47E-30 |
| | Atom2 | W | 0.170 | 5.03E-30 |
| | Atom3 | S | -0.80 | 2.53E-14 |
| $WSe_2$ | Atom1 | Se | 0.140 | 2.90E-30 |
| | Atom2 | W | -0.140 | 2.90E-30 |
| | Atom3 | Se | 0 | 9.11E-31 |
| $WTe_2$ | Atom1 | Te | 0.280 | 9.41E-29 |
|  | Atom2 | W | -0.490 | 1.34E-24 |
|  | Atom3 | Te | 0.210 | 1.24E-29 |

### 3.5.2 Reduced mass of tropical geometry calculation

By integrating tropical geometry with the reduced mass equation in Schrödinger's equation, we propose a way to re-characterize the reduced mass using tropical operations. Expressing the reduced mass $-\hbar^2\overline{M}$ and potential energy terms $E_k$ as tropical polynomials:

$$f\left(x,y\right)=\hat{E}_{\mathrm{k}}x^2+\hat{E}_{\mathrm{k}}y^2 \tag{67}$$

considering that the eigenvalues $E_k$ of the corresponding tropical geometric matrix $A=\begin{bmatrix} E_k & 0 \\ 0 & E_k \end{bmatrix}$ correspond to chemical bonds $\delta V_{bc}^{bonding}\left(\overrightarrow{r_{ij}}\right)$, we have:

$$E_k=\delta V_{bc}^{bonding}\left(\overrightarrow{r_{ij}}\right)$$

Thus, the chemical bond Mo-S correspondence matrix in the heterojunction $MoS_2/WSe_2$ is:

$$A=\begin{bmatrix} -0.5590 & 0 \\ 0 & -0.5590 \end{bmatrix}$$

The tropical geometric quadratic polynomial corresponding to this $\left(2\times2\right)$ matrix is:

$$f\left(x,y\right)=-0.5590x^2-0.5590y^2=\hbar^2\overline{M}.$$

Similarly to $MoS_2/MoTe_2$, $WSe_2/MoTe_2$, $WS_2$, $WSe_2$ and $WTe_2$, a tropical gradient map can be further constructed to display the gradient relationship between reduced mass $-\hbar^2\overline{M}$, potential energy $E_k$, and their corresponding tropical matrices $A$, as shown in **Fig. 14**.

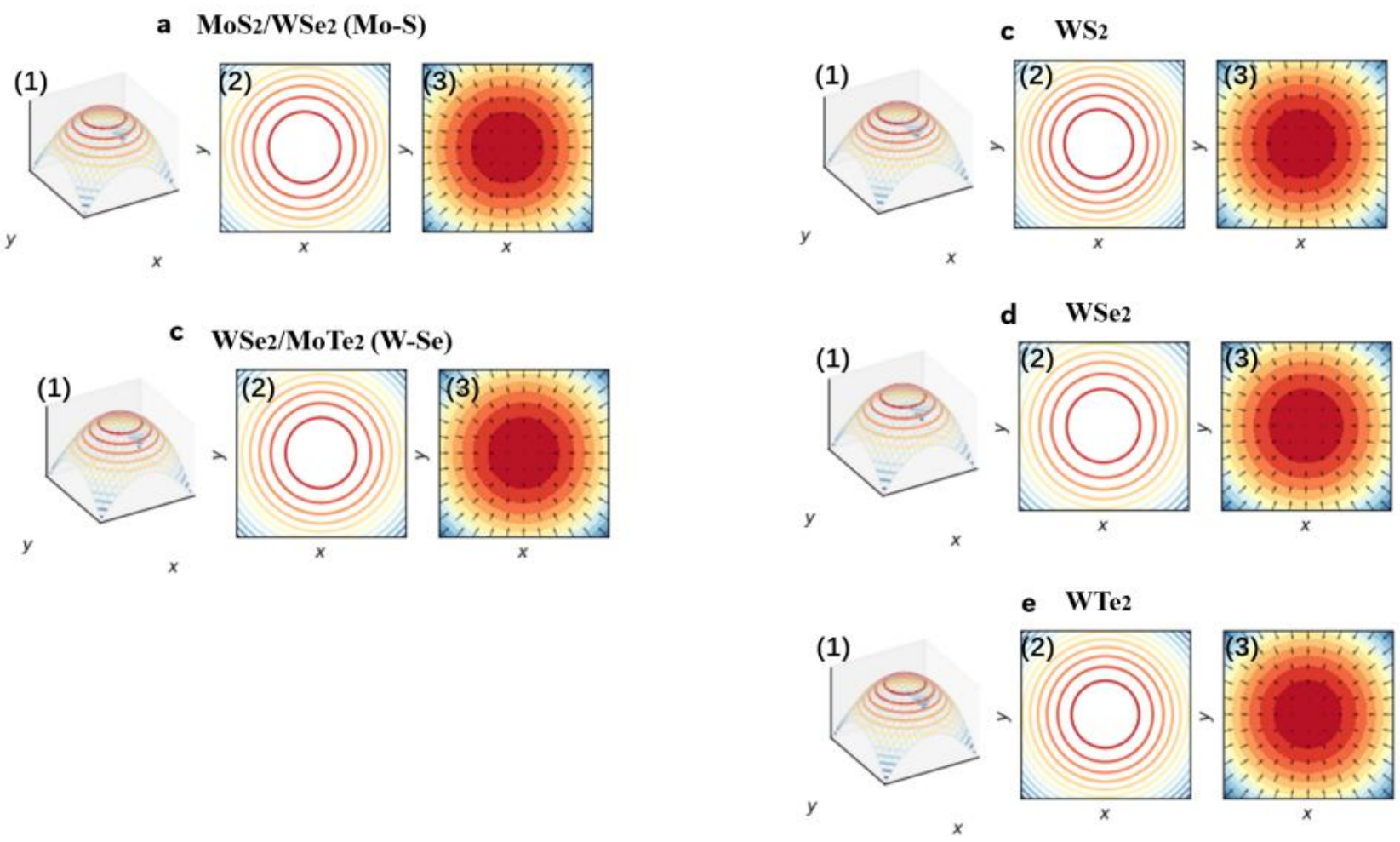

**Fig. 14** Tropical gradient map of $MoS_2/WSe_2$, $WSe_2/MoTe_2$, $WS_2$, $WSe_2$ and $WTe_2$

The 3D surface maps of the quadratic polynomials of six heterojunctions, illustrated in **Fig. 14 a**, are as follows: $f(x,y) = -0.5590x^2 - 0.5590y^2$ of Mo-S in $MoS_2/WSe_2$, $f(x,y) = -0.3569x^2 - 0.3569y^2$ of W-Se in $WSe_2/MoTe_2$, $f(x,y) = -0.3795x^2 - 0.3795y^2$ of $WS_2$, $f(x,y) = -0.3303x^2 - 0.3303y^2$ of $WSe_2$, $f(x,y) = -0.2361x^2 - 0.2361y^2$ of $WTe_2$. The 2D contour plots of the function values of the quadratic polynomial, which are reduced mass $-\hbar^2\bar{M}$ of the six heterojunctions, are shown in **Fig. 14 b**. **Fig. 14 c** depicts the vector of the gradient field based on the 2D contour map converted to reduced mass $-\hbar^2\bar{M}$.

**4. Conclusions**

This study systematically investigates the electronic structure and chemical bonding properties of two-dimensional bilayer transition metal dichalcogenides ($MX_2$, M = Mo, W; X = S, Se, Te). Using DFT in conjunction with the BBC model, we analyze the effects of atomic bonding and electronic states, focusing on the band gaps, deformation binding energies, and non-Hermitian bonding characteristics of different $MX_2$ compounds.The results indicate that charge transfer plays a significant role in reduced mass fluctuations, while tropical geometry analysis further contributes to the reduced mass. Additionally, heterostructures formed by different atomic layers effectively modulate the band gap of two-dimensional bilayer transition metal dichalcogenides. All compounds exhibit semiconductor properties, with band gaps varying within a specific range.

The electronic band structure and DOS reveal the orbital characteristics and interactions of electrons. The d orbitals of M and the p orbitals of X primarily contribute to the electronic states near the Fermi level, with orbital hybridization playing a crucial role in bond formation.This study provides a theoretical foundation for further exploration of these materials, particularly for electronic devices and optoelectronic applications. Future research may focus on optimizing heterostructures and exploring the properties of other transition metal dichalcogenide systems.

**Acknowledgment**

This research was financially supported by the Science and Technology Research Project of Fuling District, Chongqing Municipality (Grant No. FLKJ-2024BAG5128).

## References:

[1] Y. Li, F. Wu, J. Qian, M. Zhang, Y. Yuan, Y. Bai, C. Wu, Metal chalcogenides with heterostructures for high‐performance rechargeable batteries, Small Science 1 (2021) 2100012.
[2] X. Liu, F. Xu, Z. Li, Z. Liu, W. Yang, Y. Zhang, H. Fan, H.Y. Yang, Design strategy for MXene and metal chalcogenides/oxides hybrids for supercapacitors, secondary batteries and electro/photocatalysis, Coordination Chemistry Reviews 464 (2022) 214544.
[3] O. Samy, S. Zeng, M.D. Birowosuto, A. El Moutaouakil, A review on MoS2 properties, synthesis, sensing applications and challenges, Crystals 11 (2021) 355.
[4] M. Bo, H. Li, Z. Huang, L. Li, C. Yao, Bond relaxation and electronic properties of two-dimensional Sb/MoSe2 and Sb/MoTe2 van der Waals heterostructures, AIP Advances 10 (2020).
[5] Y. Tan, M. Bo, Electrostatic shielding effects and binding energy shifts and topological phases of bilayer molybdenum chalcogenides, ChemistrySelect 9 (2024) e202303817.
[6] M.U. Shahid, T. Najam, M.H. Helal, I. Hossain, S.M. El-Bahy, Z.M. El-Bahy, A. ur Rehman, S.S.A. Shah, M.A. Nazir, Transition metal chalcogenides and phosphides for photocatalytic H2 generation via water splitting: a critical review, International Journal of Hydrogen Energy 62 (2024) 1113-1138.
[7] D. Ge, R. Luo, X. Wang, L. Yang, W. Xiong, F. Wang, Internal and external electric field tunable electronic structures for photocatalytic water splitting: Janus transition-metal chalcogenides/C3N4 van der Waals heterojunctions, Applied Surface Science 566 (2021) 150639.
[8] Y. Chen, Y. Wang, Z. Wang, Y. Gu, Y. Ye, X. Chai, J. Ye, Y. Chen, R. Xie, Y. Zhou, Unipolar barrier photodetectors based on van der Waals heterostructures, Nature Electronics 4 (2021) 357-363.
[9] H. Yang, A short review on heterojunction photocatalysts: Carrier transfer behavior and photocatalytic mechanisms, Materials Research Bulletin 142 (2021) 111406.
[10] F. Li, G. Zhu, J. Jiang, L. Yang, F. Deng, X. Li, A review of updated S-scheme heterojunction photocatalysts, Journal of Materials Science & Technology 177 (2024) 142-180.
[11] Q. Xu, S. Wageh, A.A. Al-Ghamdi, X. Li, Design principle of S-scheme heterojunction photocatalyst, Journal of Materials Science & Technology 124 (2022) 171-173.
[12] S. Wang, S. Zhao, X. Guo, G. Wang, 2D material‐based heterostructures for rechargeable batteries, Advanced Energy Materials 12 (2022) 2100864.
[13] D.M. Kennes, M. Claassen, L. Xian, A. Georges, A.J. Millis, J. Hone, C.R. Dean, D. Basov, A.N. Pasupathy, A. Rubio, Moiré heterostructures as a condensed-matter quantum simulator, Nature Physics 17 (2021) 155-163.
[14] B. Huang, M.A. McGuire, A.F. May, D. Xiao, P. Jarillo-Herrero, X. Xu, Emergent phenomena and proximity effects in two-dimensional magnets and heterostructures, Nature Materials 19 (2020) 1276-1289.
[15] H. Chen, S. Teale, B. Chen, Y. Hou, L. Grater, T. Zhu, K. Bertens, S.M. Park, H.R. Atapattu, Y. Gao, Quantum-size-tuned heterostructures enable efficient and stable inverted perovskite solar cells, Nature Photonics 16 (2022) 352-358.
[16] L. Ge, M. Bo, Atomic bonding states of metal and semiconductor elements, Physica Scripta 98 (2023) 105908.
[17] J.E. Zimmermann, M. Axt, F. Mooshammer, P. Nagler, C. Schüller, T. Korn, U. Höfer, G. Mette, Ultrafast charge-transfer dynamics in twisted MoS2/WSe2 heterostructures, ACS nano 15 (2021) 14725-14731.
[18] Y. Balaji, Q. Smets, Á. Śzabo, M. Mascaro, D. Lin, I. Asselberghs, I. Radu, M. Luisier, G. Groeseneken, MoS2/MoTe2 heterostructure tunnel FETs using gated Schottky contacts, Advanced Functional Materials 30 (2020) 1905970.
[19] S. Hussain, S.A. Patil, D. Vikraman, I. Rabani, A.A. Arbab, S.H. Jeong, H.-S. Kim, H. Choi, J. Jung, Enhanced electrocatalytic properties in MoS2/MoTe2 hybrid heterostructures for dye-sensitized solar cells, Applied Surface Science 504 (2020) 144401.
[20] J.F. Sierra, J. Fabian, R.K. Kawakami, S. Roche, S.O. Valenzuela, Van der Waals heterostructures for spintronics and opto-spintronics, Nature Nanotechnology 16 (2021) 856-868.
[21] Y. Li, J. Zhang, Q. Chen, X. Xia, M. Chen, Emerging of heterostructure materials in energy storage: a review, Advanced Materials 33 (2021) 2100855.
[22] P.V. Pham, S.C. Bodepudi, K. Shehzad, Y. Liu, Y. Xu, B. Yu, X. Duan, 2D heterostructures for ubiquitous electronics and optoelectronics: principles, opportunities, and challenges, Chemical reviews 122 (2022) 6514-6613.
[23] L. Pi, P. Wang, S.-J. Liang, P. Luo, H. Wang, D. Li, Z. Li, P. Chen, X. Zhou, F. Miao, Broadband convolutional processing using band-alignment-tunable heterostructures, Nature Electronics 5 (2022)

248-254.
[24] E.M. Stein, R. Shakarchi, Complex analysis, Princeton University Press2010.
[25] R. Penrose, W. Rindler, Spinors and space-time, Cambridge university press1984.
[26] T. Needham, Visual complex analysis, Oxford University Press2023.
[27] D. Maclagan, B. Sturmfels, Introduction to tropical geometry, American Mathematical Society2021.
[28] C. Kittel, P. McEuen, Introduction to solid state physics, John Wiley & Sons2018.
[29] Y. Tan, M. Bo, Non‐Hermitian Bonding and Electronic Reconfiguration of Ba2ScNbO6 and Ba2LuNbO6, Annalen der Physik (2024) 2400040.

# Supplemental Material

Electronic Structure, mass fluctuation and Localized Bond Properties of two-dimensional double-layer transition metal chalcogenide $MX_2$ (M = Mo, W; X = S, Se, Te) Calculated Based on DFT and BBC model

Yaorui Tan and Maolin Bo*

Key Laboratory of Extraordinary Bond Engineering and Advanced Materials Technology (EBEAM) of Chongqing, Yangtze Normal University, Chongqing 408100, China

Corresponding Author: *E-mail: bmlwd@yznu.edu.cn (Maolin Bo)

## 1. Structural stability

We have calculated the phonon spectra of $MoS_2/WSe_2$, $WSe_2/MoTe_2$, $WS_2$, $WSe_2$, and $WTe_2$. The results show that there are no negative frequencies in these materials, indicating that they have stable structures.

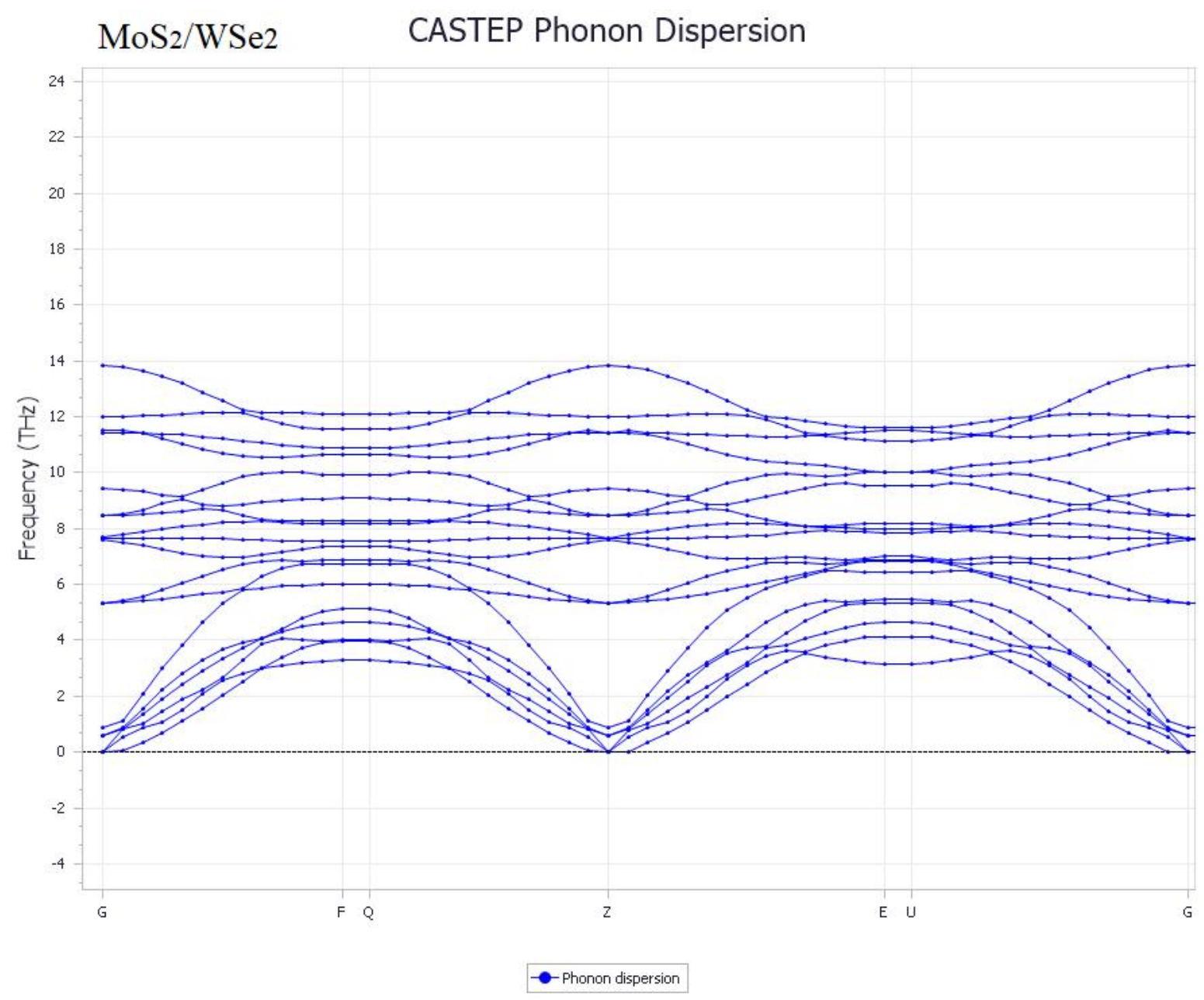

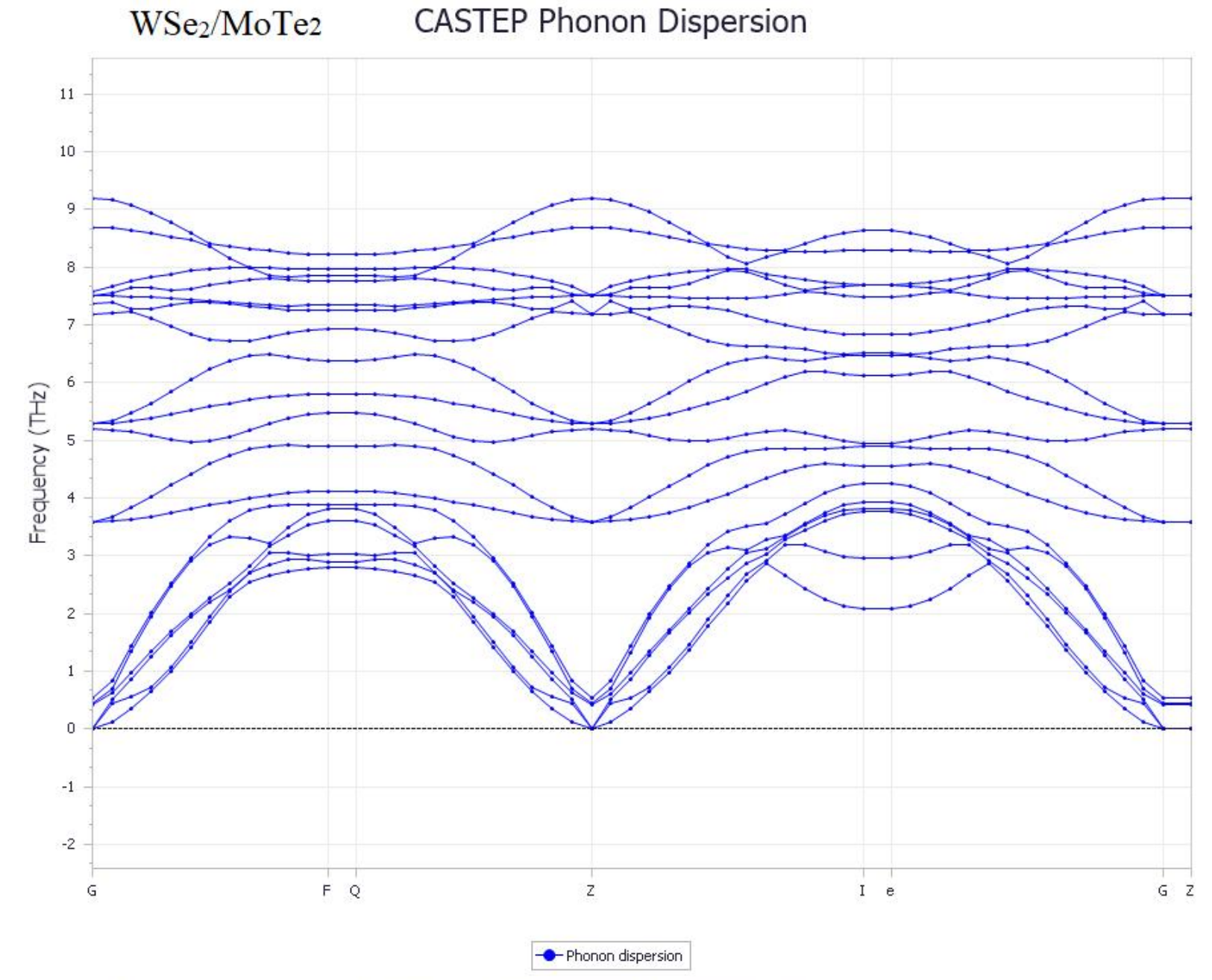

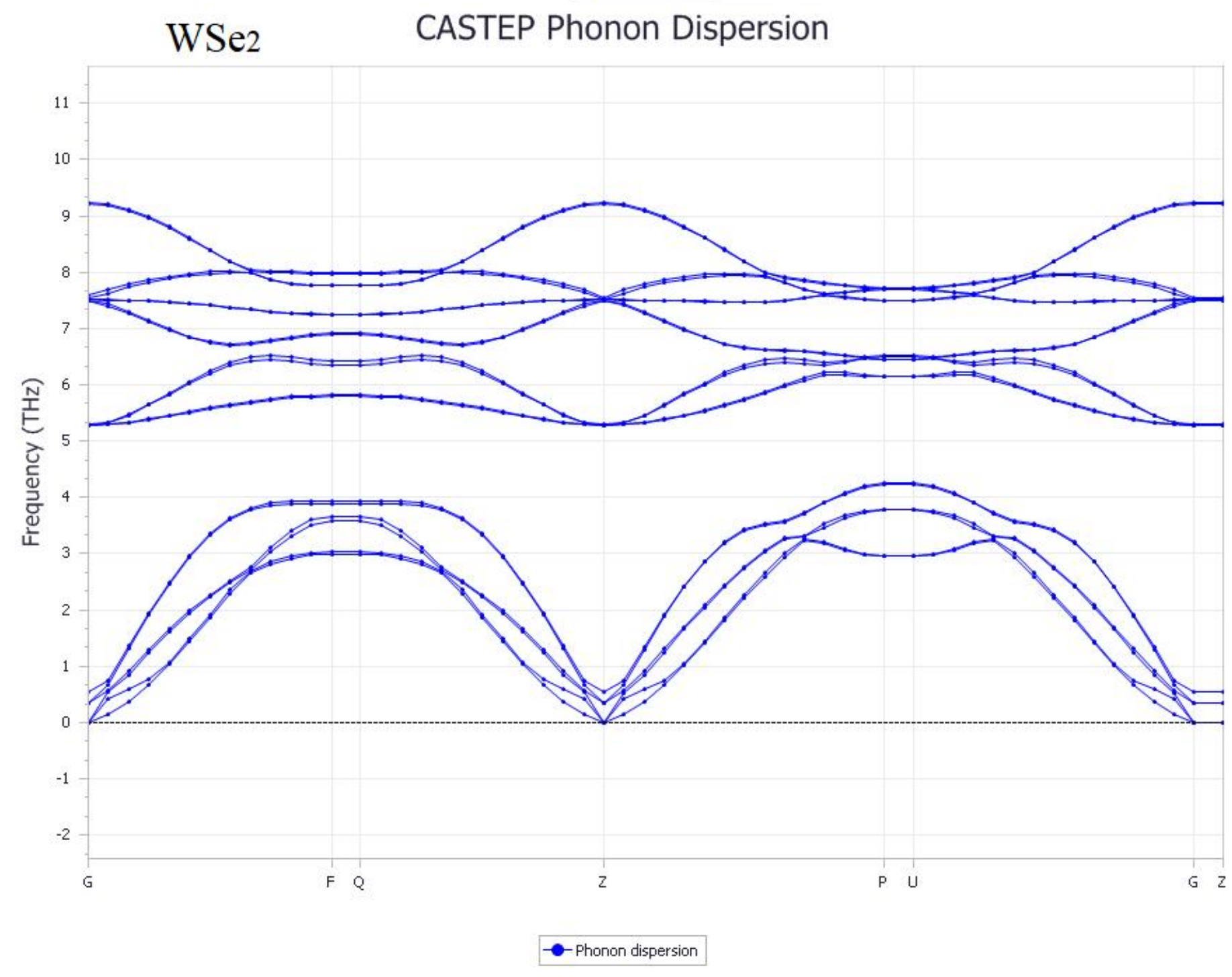

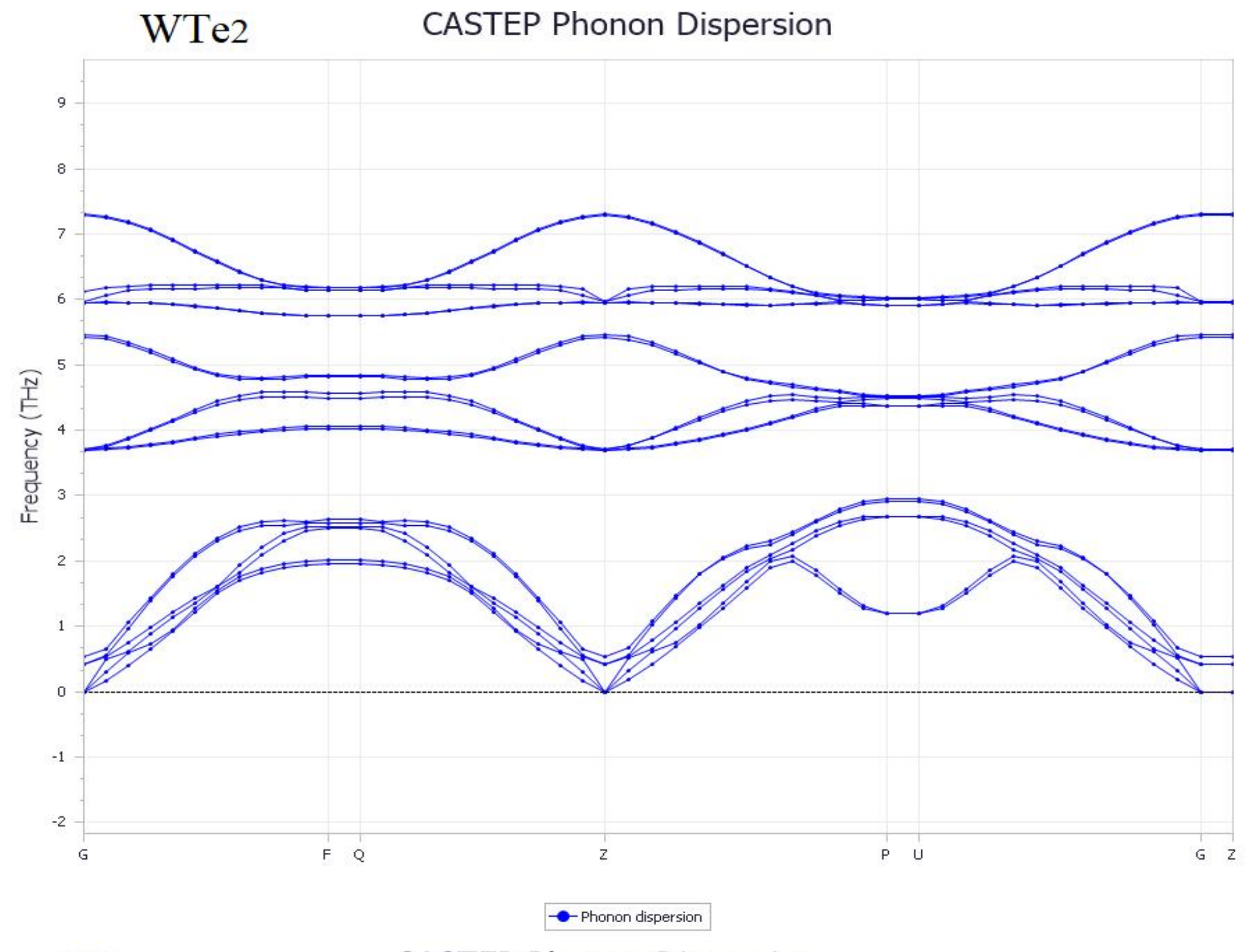

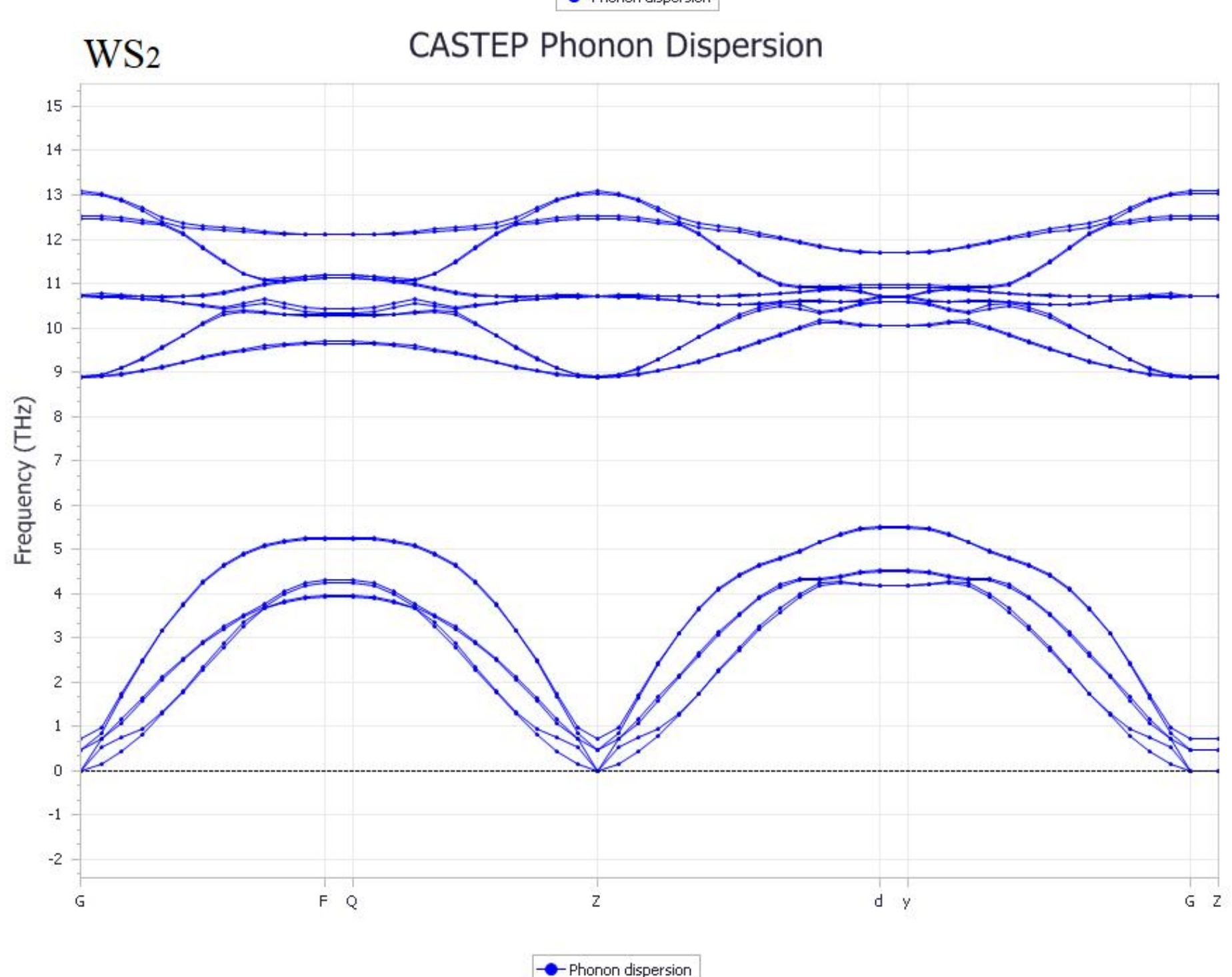

## 2. Lorentz transformation

According to special relativity, the laws of physics hold equally in a reference frame moving at a constant velocity as they do in a stationary frame. One intriguing consequence of this principle is that it is impossible to distinguish which system (if any) is truly stationary, making it impossible to determine the exact speed of any other system.

In special relativity, the laws of physics remain valid in all inertial frames. An inertial frame is a reference system that adheres to Newton's first law of inertia, meaning an object will maintain constant velocity in a straight line unless acted upon by an external force. Since any two inertial frames must move at a constant velocity relative to each other, any frame that moves at a constant speed relative to an existing inertial frame is also considered an inertial system.

Assuming we have two inertial frames *S* and *S′* moving relative to *S* at a constant velocity *V* (magnitude *V*) (*S* therefore moves relative to *S′* at a velocity - *V*). We can also take the coordinates $x/x'$ so that the motion follows the usual axis (see Figure 1), and place the clocks of each inertial frame at the origin, so that both clocks read zero when the two frames coincide (i.e. $x = x' = 0$, at that time $t = t' = 0$). Assuming that an event $(x, y, z)$ occurs at time *t* in the S system.

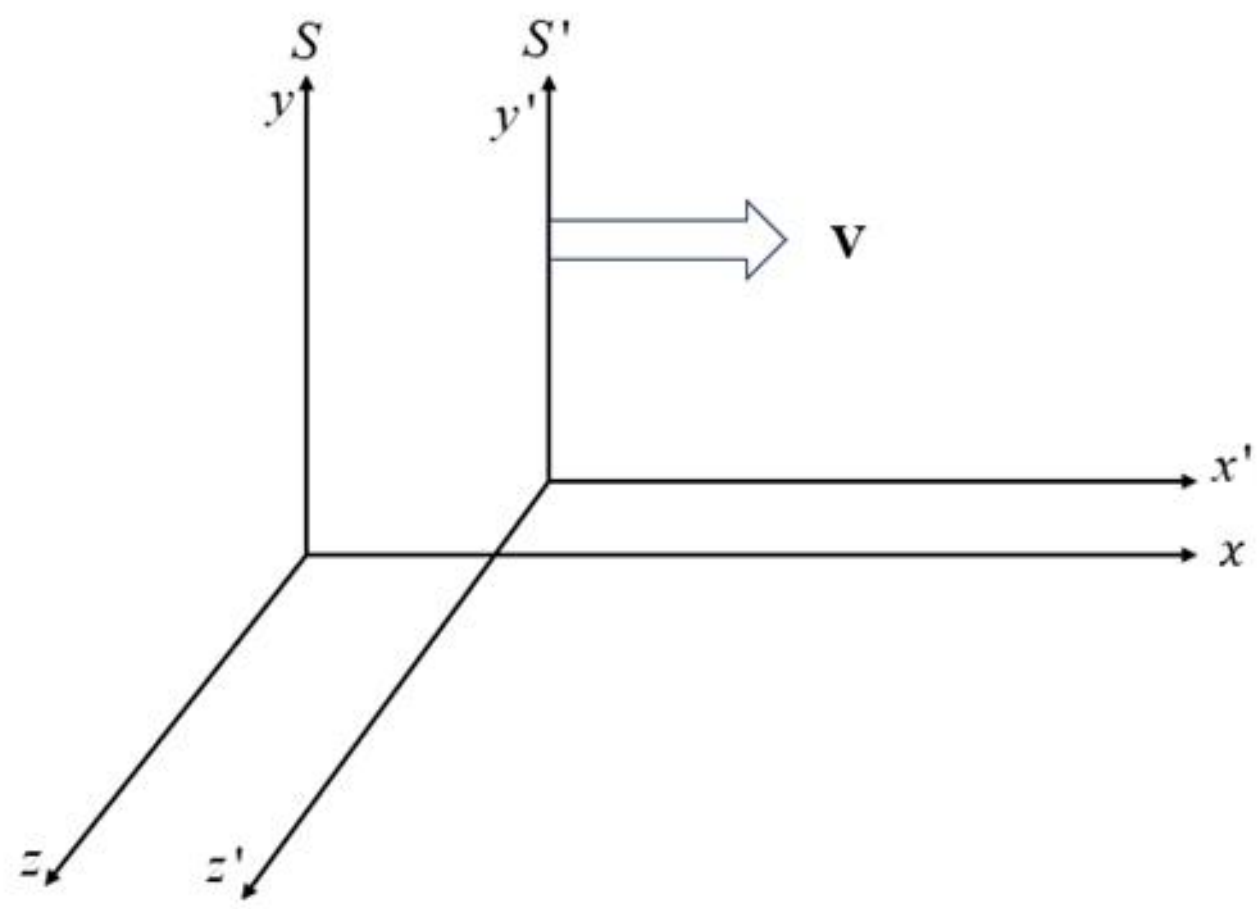

**Figure 1**

What are the spatiotemporal coordinates $t'$ and $(x', y', z')$ of the same event in the *S′* system? The answer is provided by the Lorentz transformation:

$$x' = \gamma(x - vt)$$
$$y' = y$$
$$z' = z$$
$$t' = \gamma\left(t - \frac{v}{c^2}x\right)$$

among

$$\gamma \equiv \frac{1}{\sqrt{1 - \frac{v^2}{c^2}}}$$

The inverse exchange from $S'$ to $S$ can be easily changed. The symbol obtained is:

$$x = \gamma(x' - vt')$$
$$y = y'$$
$$z = z'$$
$$t = \gamma\left(t' - \frac{v}{c^2}x'\right)$$

**3. Four vectors**

We define the position time four vectors $x^\mu$ , $\mu = 0,1,2,3$ as follows:

$$x^0 = ct, x^1 = x, x^2 = y, x^3 = z$$

The Lorentz transformation can be written in a more symmetrical form:

$$x^{0'} = \gamma\left(x^0 - \beta x^1\right)$$
$$x^{1'} = \gamma\left(x^1 - \beta x^0\right)$$
$$x^{2'} = x^2$$
$$x^{3'} = x^3$$

among

$$\beta \equiv \frac{v}{c}$$

Tighter and more compact

$$x^{\mu'} = \sum_{v=0}^{3} \Lambda_v^\mu x^v, (\mu = 0,1,2,3)$$

The coefficient $\Lambda_v^\mu$ can be seen as a matrix element of a matrix $\Lambda$:

$$\Lambda = \begin{pmatrix} \gamma & -\gamma\beta & 0 & 0 \\ -\gamma\beta & \gamma & 0 & 0 \\ 0 & 0 & 1 & 0 \\ 0 & 0 & 0 & 1 \end{pmatrix}$$

(i.e $\Lambda_0^0 = \Lambda_1^1 = \gamma; \Lambda_0^1 = \Lambda_1^0 = -\gamma\beta; \Lambda_2^2 = \Lambda_3^3 = 1.$; all others are 0). To simplify notation and avoid excessive summation signs, we adopt Einstein's summation rule, which implies automatic summation over repeated Greek indices (one subscript, one superscript) from 0 to 3. Therefore, the equation is written as follows at the end:

$$x^{\mu'} = \Lambda_\nu^\mu x^\nu$$

The special advantage of this neat expression is that it has the same form for Lorentz transformations that are not along the *x*-direction. In fact, the coordinate axes of *S* and *S′* do not even need to be parallel to each other. The matrix $\Lambda$ is naturally more complex, but equation still holds.

When transitioning from *S* to *S′*, the individual coordinates of the event may change, but a special combination of them remains unchanged:

$$I = \left(x^0\right)^2 - \left(x^1\right)^2 - \left(x^2\right)^2 - \left(x^3\right)^2 = \left(x^{0'}\right)^2 - \left(x^{1'}\right)^2 - \left(x^{2'}\right)^2 - \left(x^{3'}\right)^2$$

The quantity that has the same value in any inertial frame is called an invariant. Similarly, if an appropriate angle $\theta$ is chosen as the pivot (counterclockwise pivot):

$$\begin{cases} x \to x\cos\theta + y\sin\theta \\ y \to -x\sin\theta + y\cos\theta \end{cases}$$

Represent by matrix

$$\begin{pmatrix} x \\ y \end{pmatrix} \to \begin{pmatrix} x\cos\theta + y\sin\theta \\ -x\sin\theta + y\cos\theta \end{pmatrix} = \begin{pmatrix} \cos\theta & \sin\theta \\ -\sin\theta & \cos\theta \end{pmatrix} \begin{pmatrix} x \\ y \end{pmatrix}$$

Represent by matrix components

$$x'_i = R_i^j x_j, \quad x_i = \begin{pmatrix} x \\ y \end{pmatrix}, \quad i, j = 1, 2$$

Take transpose

$$\begin{pmatrix} x' & y' \end{pmatrix} = \begin{pmatrix} x & y \end{pmatrix} \begin{pmatrix} \cos\theta & -\sin\theta \\ \sin\theta & \cos\theta \end{pmatrix}$$

$$x'^i = x^j \left(R^T\right)_j^i, \quad x^j = \begin{pmatrix} x & y \end{pmatrix}, \quad i, j = 1, 2, \quad R_i^j x_j = R_i^1 x_1 + R_i^2 x_2$$

therefore

$$R^T R = \begin{pmatrix} \cos\theta & -\sin\theta \\ \sin\theta & \cos\theta \end{pmatrix} \begin{pmatrix} \cos\theta & \sin\theta \\ -\sin\theta & \cos\theta \end{pmatrix}$$

$$= \begin{pmatrix} \cos^2\theta + \sin^2\theta & \cos\theta\sin\theta - \sin\theta\cos\theta \\ \sin\theta\cos\theta - \cos\theta\sin\theta & \cos^2\theta + \sin^2\theta \end{pmatrix}$$

$$= \begin{pmatrix} 1 & 0 \\ 0 & 1 \end{pmatrix} = I$$

So $r^2 = x^2 + y^2 + z^2$ remains constant under rotation.

Now, we express this invariant in summation form: $\sum_{\mu=0}^{3} x^\mu x^\mu$ , However, this expression contains three repeated negative signs in the summation. To retain their traces, we introduce a metric $g_{\mu\nu}$ whose components can be represented as a matrix $g$:

$$g = \begin{pmatrix} 1 & 0 & 0 & 0 \\ 0 & -1 & 0 & 0 \\ 0 & 0 & -1 & 0 \\ 0 & 0 & 0 & -1 \end{pmatrix}$$

(i.e. $g_{00} = 1; g_{11} = g_{22} = g_{33} = -1$, all others are 0). With the help of $g_{\mu\nu}$ , invariants $I$ can be written in the form of double sums:

$$I = \sum_{\mu=0}^{3} \sum_{\nu=0}^{3} g_{\mu\nu} x^\mu x^\nu = g_{\mu\nu} x^\mu x^\nu$$

Furthermore, we define covariant four vectors $x_\mu$ (with indicators listed below) as follows:

$$x_\mu \equiv g_{\mu\nu} x^\nu$$

(That is, $x_0 = x^0, x_1 = -x^1, x_2 = -x^2, x_3 = -x^3$). To emphasize the difference, we call the "original" four vector $x^\mu$ (with the index above) the inverted four vector.

Invariants can therefore be written in the simplest form:

$$I = x^\mu x_\mu$$

(Or equivalently written $x_\mu x^\mu$ ). Once familiar with this method, it becomes straightforward. (Furthermore, it effectively extends non-Cartesian coordinate systems to the curved spaces encountered in general relativity, although neither of these things is relevant to us here.)

Bit-time four vectors $x^\mu$ are typical of all four vectors. When we change from one inertial space to another, we define four vectors $a^{\mu}$ as a four component quantity to transform in the same way as $x^{\mu}$, that is:

$$a^{\mu'} = \Lambda^{\mu}_{v} a^{v}$$

Using the same coefficient $\Lambda^{\mu}_{v}$ . For each of these (inverse) four vectors, we introduce a covariant four vector $a_{\mu}$ by changing the sign of its spatial component

$$a_{\mu} = g_{\mu v} a^{v}$$

Of course, we can also reverse from co empty to inverse by changing the symbol again:

$$a^{\mu} = g_{\mu v} a_{v}$$

Technically, this $g^{\mu v}$ is the element of a matrix $g^{-1}$ (however, since our metric is its own inverse, $g^{\mu v}$ and $g_{\mu v}$ are the same). Given any two four vectors, $a^{\mu}$ and $b^{\mu}$ :

$$a^{\mu} b_{\mu} = a_{\mu} b^{\mu} = a^{0} b^{0} - a^{1} b^{1} - a^{2} b^{2} - a^{3} b^{3}$$

It is invariant (with the same value in any inertial frame). We will call it the scalar product of *a* and *b*: this is the four-dimensional analog corresponding to the dot product of two or three vectors (a four-dimensional analog without cross product).

If you don't want to write subscripts, you can use dot product instead. Position vector dot product:

$$\vec{x} \cdot \vec{x} = x^{i} x_{i} = \delta_{ij} x^{i} x^{j} = \begin{pmatrix} x & y \end{pmatrix} \begin{pmatrix} x \\ y \end{pmatrix} = x^{2} + y^{2}$$

Denoted $\Lambda$ as Lorentz transformation matrix:

$$S^{2} = x^{\mu} x_{\mu} = g_{\mu v} x^{\mu} x^{v} = g_{\rho\sigma} x^{\rho} x^{\sigma}$$

$$S'^{2} = x'^{\mu} x'_{\mu} = g_{\mu v} x'^{\mu} x'^{v} = g_{\mu v} \Lambda^{\mu}_{\rho} x^{\rho} \Lambda^{v}_{\sigma} x^{\sigma}$$

The transformation that satisfies the following conditions is called Lorentz transformation:

$$S^{2} = S'^{2} \Rightarrow g_{\rho\sigma} = g_{\mu v} \Lambda^{\mu}_{\rho} \Lambda^{v}_{\sigma}$$

So, it can be written as:

$$g_{\rho\sigma} = g_{\mu v} \Lambda^{\mu}_{\rho} \Lambda^{v}_{\sigma} = \left(\Lambda^{T}\right)^{\mu}_{\rho} g_{\mu v} \Lambda^{v}_{\sigma}$$

It can be written in the form of a matrix:

$$\Lambda^T g \Lambda = g = \begin{pmatrix} -1 & & & \\ & -1 & & \\ & & -1 & \\ & & & -1 \end{pmatrix}$$

The transformation matrix form is as follows:

$$\begin{pmatrix} 1 & & & \\ & 1 & & \\ & & \cos\theta_x & -\sin\theta_x \\ & & \sin\theta_x & \cos\theta_x \end{pmatrix}, \quad \begin{pmatrix} 1 & & & \\ & \cos\theta_y & & \sin\theta_y \\ & & 1 & \\ & -\sin\theta_y & & \cos\theta_y \end{pmatrix}$$

$$\begin{pmatrix} 1 & & & \\ & \cos\theta_z & -\sin\theta_z & \\ & -\sin\theta_z & \cos\theta_z & \\ & & & 1 \end{pmatrix}, \quad \begin{pmatrix} \cosh Y_x & -\sinh Y_x & & \\ -\sinh Y_x & \cosh Y_x & & \\ & & 1 & \\ & & & 1 \end{pmatrix}$$

$$\begin{pmatrix} \cosh Y_y & & -\sinh Y_y & \\ & 1 & & \\ -\sinh Y_y & & \cosh Y_y & \\ & & & 1 \end{pmatrix}, \quad \begin{pmatrix} \cosh Y_z & & & -\sinh Y_z \\ & 1 & & \\ & & 1 & \\ -\sinh Y_z & & & \cosh Y_z \end{pmatrix}$$

therefore

$$\cosh Y_x = \frac{1}{\sqrt{1-v^2}}, \quad \sinh Y_x = \frac{v}{\sqrt{1-v^2}}$$

Obtain transformation:

$$x \to \Lambda^{-1} x = \frac{x+vt}{\sqrt{1-v^2}}, \quad t \to \Lambda^{-1} t = \frac{t+vx}{\sqrt{1-v^2}},$$

Equivalent to:

$$x \to x+vt, t \to t'$$

The Lorentz inverse transform satisfies the form:

$$x'^{\mu} = \Lambda^{\mu}_{\nu} x^{\nu}, \quad x^{\nu} = \left(\Lambda^{-1}\right)^{\mu}_{\gamma} x'^{\gamma}$$

Orthogonality

$$x'^{\mu} = \Lambda^{\mu}_{\nu} x^{\nu} = \Lambda^{\mu}_{\nu} \left(\Lambda^{-1}\right)^{\mu}_{\gamma} x'^{\gamma} = \delta^{\mu}_{\gamma} x'^{\gamma}, \Lambda^{\mu}_{\nu} \left(\Lambda^{-1}\right)^{\nu}_{\gamma} = \delta^{\mu}_{\gamma}$$

$$g_{\rho\sigma} \left(\Lambda^{-1}\right)^{\rho}_{\tau} = g_{\mu\nu} \Lambda^{\mu}_{\rho} \Lambda^{\nu}_{\sigma} \left(\Lambda^{-1}\right)^{\rho}_{\tau} = g_{\mu\nu} \Lambda^{\nu}_{\sigma} \delta^{\mu}_{\tau} = g_{\tau\nu} \Lambda^{\nu}_{\sigma}$$

$$g^{\gamma\sigma}g_{\gamma\sigma}\left(\Lambda^{-1}\right)_{\tau}^{\rho}=\delta_{\rho}^{\tau}\left(\Lambda^{-1}\right)_{\tau}^{\rho}=g^{\gamma\sigma}g_{\gamma\sigma}\Lambda_{\sigma}^{v}\Rightarrow\left(\Lambda^{-1}\right)_{\tau}^{\gamma}g^{\gamma\sigma}g_{\gamma\sigma}\Lambda_{\sigma}^{v}$$

$$x'^{\mu}=g_{\mu v}x'_{v}=g^{\mu v}x_{\rho}\Lambda_{v}^{\rho}=g^{\mu v}\Lambda_{v}^{\rho}g_{\mu v}x^{\sigma}=\left(\Lambda^{-1}\right)_{\sigma}^{\mu}x^{\sigma},\phi(x')\rightarrow\phi(x)$$

$$V^{\mu}(x)\rightarrow\Lambda_{v}^{\mu}V^{v}(x)=\Lambda_{v}^{\mu}V^{v}\left(\Lambda^{-1}x\right)$$

Classify all four vectors by symbol:

$$\Lambda^{\mu}V_{\mu}\begin{cases}>0, timelike\\ <0, spacelike\\ =0, lightlike\end{cases}$$

Starting from vectors, we arrive at tensors in a short step: a second-order tensor $s^{\mu v}$ carrying two indicators, there are $4^2=16$ components, and two transformations $\Lambda$:

$$s^{\mu v'}=\Lambda_{k}^{\mu}\Lambda_{\sigma}^{v}s^{k\sigma}$$

A third-order tensor $t^{\mu v\lambda}$ carries three indicators, there are $4^3=64$ components, and is transformed using three transformations $\Lambda$:

$$t^{\mu v\lambda'}=\Lambda_{k}^{\mu}\Lambda_{\sigma}^{v}\Lambda_{\tau}^{\lambda}t^{k\sigma\tau}$$

In this hierarchy, vectors are tensors of order 1, and scalars (invariant) are tensors of order 0. We construct covariant and "mixed" tensors by reducing the indicators (at the cost of adding a negative sign to each spatial indicator), for example

$$s_{v}^{\mu}=g_{v\lambda}s^{\mu\lambda};s_{\mu v}=g_{\mu k}g_{v\lambda}s^{k\lambda}$$

Note that the product of two tensors itself is also a tensor: ($a^{\mu}b^{v}$) is a second-order tensor; ($a^{\mu}t^{v\lambda\sigma}$) is a fourth-order tensor; wait. Finally, we can "shrink" any *n*+2 order tensor into an n-order tensor by summing a pair of upper and lower indices. Therefore, $s_{\mu}^{\mu}$ is a scalar; $t_{v}^{\mu v}$ is a vector; $a_{\mu}t^{\mu v\lambda}$ is a second-order tensor.

According to relativity, a moving clock (such as your watch) ticks slower compared to a stationary clock on the ground. Similarly, bodily functions and mental processes slow down when observed from the ground frame. Specifically, when the ground clock measures an infinitesimal time interval $dt$ $dt$, your own (or proper) time experiences a small amount $d\tau$:

$$d\tau=\frac{dt}{\gamma} \tag{1}$$

Of course, at usual speeds, $\gamma$ is close to 1, making $dt$ and $d\tau$ almost equal. However, in elementary particle physics, the difference between laboratory time (the reading of the wall clock) and particle time (as if there were a watch on the particle) is crucial. Although Equation (1) allows us to convert one time measure into the other, in practice, using proper time is often more convenient. This is because dτd\taudτ is an invariant, meaning that all observers can agree on the particle's own time, even if their individual clocks differ from it and from one another.

When referring to the velocity of a particle relative to the laboratory frame, we define it as the distance traveled (measured in the laboratory frame) divided by the time taken (measured by the laboratory clock):

$$v = \frac{dx}{dt} \tag{2}$$

However, it is useful to introduce the proper velocity η, which is defined as the distance traveled (also measured in the laboratory frame) divided by the proper time:

$$\eta = \frac{dx}{d\tau} \tag{3}$$

According to equation (3.29), the two velocities are related by the factor $\gamma$:

$$\eta = \gamma v \tag{4}$$

$\eta$ is easier to handle because when transforming from the laboratory frame $S$ to the moving frame $S'$, both the numerator and the denominator in equation (2) must be transformed—leading to the cumbersome velocity addition rule in equation—while in equation (3), only the numerator needs to be transformed; $d\tau$ is invariant. In fact, the proper velocity is part of a four-vector:

$$\eta^{\mu} = \frac{dx^{\mu}}{d\tau} \tag{5}$$

its zero component is

$$\eta^{0} = \frac{dx^{0}}{d\tau} = \frac{d(ct)}{(1/\gamma)dt} = \gamma c \tag{6}$$

thus,

$$\eta^{\mu} = \gamma\left(c, v_x, v_y, v_z\right)$$

(7)

Incidentally, $\eta_{\mu}\eta^{\mu}$ should be invariant, and it is

$$\eta_{\mu}\eta^{\mu} = \gamma^2\left(c^2 - v_x^2 - v_y^2 - v_z^2\right) = \gamma^2 c^2\left(1 - v^2/c^2\right) = c^2$$

(8)

they do not create more invariants.

Classically, momentum is defined as $p = mv$ (p = mv). However, in the relativistic case, we must choose between ordinary velocity and proper velocity. Since these two velocities are equal in the non-relativistic limit, classical intuition does not provide guidance. Defining momentum as $p = mv$ (p = mv) would violate the consistency of momentum conservation with the principle of relativity across inertial frames. In contrast, defining momentum as $p = m\eta$ (p = mη) preserves this consistency. While experiments ultimately determine whether momentum is conserved, the relativistic generalization shows that using proper velocity $\eta$ (η) is the more suitable choice.

Thus, in relativity, momentum is defined as mass times proper velocity:

$$p = m\eta$$

(9)

since proper velocity is part of a four-vector, momentum is also

$$p^{\mu} = m\eta^{\mu}$$

(10)

the spatial components of $p$ constitute the (relativistic) momentum three-vector:

$$p = \gamma m v = \frac{mv}{\sqrt{1 - v^2/c^2}}$$

(11)

at the same time, the "time" component is

$$p^0 = \gamma m c$$

(12)

For reasons that will soon become apparent, we define the relativistic energy $E$ as

$$E \equiv \gamma m c^2 = \frac{mc^2}{\sqrt{1 - v^2/c^2}}$$

(13)

The 0-component of $p^{\mu}$ is therefore $E/c$. Thus, energy and momentum together form a four-vector—the energy-momentum four-vector (or four-momentum)

$$p^{\mu}=\left(\frac{E}{c},p_x,p_y,p_z\right)$$

(14)

incidentally, from equations (8) and (10), we have

$$p_{\mu}p^{\mu}=\frac{E^2}{c^2}-p^2=m^2c^2$$

(15)

which is also completely invariant.

The relativistic momentum [equation (9)] reduces to the classical expression in the non-relativistic region $(v \ll c)$, but the same cannot be said for the relativistic energy [equation (13)]. To understand why this quantity is referred to as "energy," we expand the square root in a Taylor series:

$$E=mc^2\left(1+\frac{1}{2}\frac{v^2}{c^2}+\frac{3}{8}\frac{v^4}{c^4}+\cdots\right)=mc^2+\frac{1}{2}mv^2+\frac{3}{8}m\frac{v^4}{c^2}+\cdots$$

(16)

note that the second term here is precisely the classical kinetic energy term, and the leading term $mc^2$ is a constant. In this sense, the relativistic formula is consistent with the classical one when the higher-order terms in the expansion at the limit $v \ll c$ can be neglected. The constant term left over at $v=0$ is called the rest energy:

$$R\equiv mc^2$$

(17)

the remaining part is the energy of a moving particle called relativistic kinetic energy:

$$T\equiv mc^2\left(\gamma-1\right)=\frac{1}{2}mv^2+\frac{3}{8}m\frac{v^4}{c^2}+\cdots$$

(18)

In classical mechanics, there are no massless particles; their momentum (*mv*) would be zero, and their kinetic energy $\frac{1}{2}mv^2$ would be zero because *F=ma* they would not be subject to any force . At first glance, the same might seem true in relativity, but a careful look at the formulae

$$p = \frac{mv}{\sqrt{1 - v^2/c^2}}, E = \frac{mc^2}{\sqrt{1 - v^2/c^2}}$$

(19)

when *m=0*, the numerator is zero; but if *v=c*, the denominator is also zero, and these equations become indeterminate (*0/0*). Therefore, we might allow *m=0*, provided that the particle always moves at the speed of light. In this case, equation (19) would not serve as the definition of *E* and *p*; however, equation (15) still holds:

$$v = c\ , \quad E = |p|c\left( for\, massless\ particles \right)$$

(20)

if it weren't for the existence of massless particles (photons) with known facts in nature, which do indeed travel at the speed of light and their energy and momentum are indeed linked through equation (20). If equation (19) cannot define *p* and *E*, quantum mechanics provides an answer in the form of Planck's formula:

$$E = hv$$

(21)

## 4. Tight-binding Approximation and Second Quantization Representation

### 4.1 Tight-binding approximation

Starting from the Wannier function, the tight-binding approximation energy band formula can be derived. When the atomic spacing in the crystal is sufficiently large, it is reasonable to approximate the Wannier function $a_n(r-l)$ with the atomic orbital function $\phi_n(r-l)$ on the lattice point *l*. at this time, the tight binding approximation energy band electron wave function

$$\psi_{nk}(r) = \frac{1}{\sqrt{N}} \sum_l e^{ik \cdot l} \phi_n(r-l)$$

(22)

assuming that the atomic orbital functions on different lattice points are approximately orthogonal:

$$\int \phi_n^*(r-l) \phi_n(r-l') dr \approx \delta u'$$

(23)

using equation $\int a_n^*(r-l)Ha_{n'}(r-l')dr = \delta_{nn'}\varepsilon_n(l'-l)$, the matrix element of the wanier function of $H$ is approximated as:

$$\begin{aligned} H_{nn}(l,l') &= \int \phi_n^*(r)H\phi_n(r-l')dr \\ &= \varepsilon_n(l'-l) \end{aligned} \tag{24}$$

substituting into $E(k) = \sum_l e^{-ik\cdot l}\varepsilon_n(l)$ and calculating up to the nearest neighbors of $\varepsilon_n(l'-l)$, we obtain:

$$\begin{aligned} E_n(k) &= \sum_{(l'-l)} e^{ik\cdot(l'-l)}\varepsilon_n(l'-l) \\ &= \varepsilon_n(0) + \sum_\rho \varepsilon_n(\rho)e^{ik\cdot\rho} \end{aligned} \tag{25}$$

$\rho$ represents the vector connecting the nearest neighbor lattice points $(l'-l)$, and the summation over $\rho$ includes Z vectors, where Z is the coordination number of the lattice. Since the atomic orbital functions are known and satisfy the equation:

$$\left.\begin{aligned} H_0\phi_n(r-l) &= E_n^a\phi_n(r-l) \\ H_0 &= -\frac{\hbar^2}{2m}\nabla^2 + \upsilon_a(r-l) \end{aligned}\right\} \tag{26}$$

where $E_n^a$ is the atomic energy level, which is also a known quantity, both $\varepsilon_n(0)$ and $\varepsilon_n(\rho)$ can be specifically calculated. Applying equations (26) and (23), we get:

$$\begin{aligned} \varepsilon_n(0) &= \int \phi_n^*(r)H\phi_n(r-l')dr \\ &= E_n^a - A_n \end{aligned} \tag{27}$$

where

$$A_n = -\int \phi_n^*(r)\left[V(r) - \upsilon_a(r-l)\right]\phi_n(r-l)dr$$

represents the energy level shift of $E_n^a$ caused by the potential of $N-1$ atoms outside the $l$ lattice point in the lattice. Similarly

$$\varepsilon_n(\rho)=\int\phi_n^*(r)H\phi_n(r-\rho)dr$$
$$=\int\phi_n^*(r)\left[V(r)-\upsilon_a(r)\right]\phi_n(r-\rho)dr$$
$$=-J_n$$

(28)

$\varepsilon_n(\rho)$ is negative, indicating that the potential fields of $(N-1)$ other atoms will transfer the bound electrons on the $l$ lattice point to the nearest neighbor lattice point. For the sake of symmetry, let $\varepsilon_n(\rho)$ of *Z* nearest neighbor lattice points be the same and denoted by $-J_n$, $J_n$ is called the overlap integral. Thus, the well-known tight binding approximate energy band formula is obtained

$$E_n(k)=E_n^a-A_n-J_n\sum_\rho e^{ik\cdot\rho}$$

(29)

when the lattice has a center of symmetry, the contribution of a pair of lattice points with opposite orientations in the summation term of $\rho$ is $-2J_n\cos(k\cdot\rho)$. Since the cosine term is between -1 and+1, the half width of the tight binding approximation energy band described in equation (9) is $ZJ_n$, the band center is at $E_n^a-A_n$, and the band index is marked by the quantum number of atomic energy levels, such as 3*s*, 3*p* and 3*d* energy bands. In the specific calculation, the periodic field $V(r)$ is represented by the superposition of atomic potential field $\upsilon_a(r-l)$

$$V(r)=\sum_l\upsilon_a(r-l)$$

(30)

the overlap integral is then written as:

$$J_n=-\sum_{l'\neq0}\int\phi_n^*(r)\upsilon_a(r-l')\phi_n(r-\rho)dr$$

(31)

Here, both two-center and three-center integrals are involved. Additionally, it important to note that the linear combination of atomic orbitals assumed in equation (22) is only valid for non-degenerate atomic energy levels.

For the *p* and *d* orbital electrons of an atom, the system falls into the degenerate case of atomic energy levels. In such cases, the tight-binding approximation wave function must be extended. To account for this, the following trial functions for a given *k* are constructed by incorporating the linear combination of each degenerate orbital $\phi_n$

$$\psi_k(r) = \sum_m C_m \phi_{mk}(r) = \frac{1}{\sqrt{N}} \sum_l \sum_m e^{ik\cdot l} C_m \phi_m(r-l) \tag{32}$$

using $C_m$ as variational parameters, for the single-electron energy, we have:

$$\left.\begin{aligned} & E(k) = \langle \psi_k | H | \psi_k \rangle = \sum_m \sum_{m'} C_m^* C_{m'} \langle \psi_{mk} | H | \psi_{mk} \rangle \\ & \sum_m |C_m|^2 = 1 \end{aligned}\right\} \tag{33}$$

by finding the variational extremum, the secular equation for determining the energy bands can be obtained:

$$\det \left\| H_{mm'}(k') - \delta_{mm'} E(k) \right\| = 0 \tag{34}$$

where

$$H_{mm'}(k') = \langle \psi_{mk} | H | \psi_{m'k} \rangle = \frac{1}{N} \sum_l \sum_{l'} e^{ik\cdot(l-l')} H_{mm'}(l,l') \tag{35}$$

$$H_{mm'}(l,l') = \int \phi_m^*(r-l) H \phi_{m'}(r-l') dr \tag{36}$$

unlike the H matrix element formula $\int a_n^*(r-l) H a_{n'}(r-l') dr = \delta_{nn'} \varepsilon_n (l'-l)$ of the wanier function, the matrix element formula (36) is off diagonal for $m$ because the $m$ and $m'$ orbitals are degenerate. For the $p$ orbitals, $\phi_m$ should be calculated as $xf(r)$, $yf(r)$, and $zf(r)$ degenerate states, and for $d$ orbitals, it involves the determinant composed of the five degenerate states $xyf(r)$, $xzf(r)$, $yzf(r)$, $(3x^2-r^2)f(r)$ and $(x^2-y^2)f(r)$.

If the orbital hybridization effect is further considered, the summation over $m$ in equation (32) must also include the interacting atomic orbitals, At this time, $\psi_k$ is extended to the trial function of atomic orbital superposition with different symmetries, such as the superposition of and $p$ orbitals.

The key principle of the tight-binding approximation method is the linear combination of atomic orbital functions, which is widely known in the literature as the linear combination of atomic orbitals (LCAO) method. The primary challenge of this method lies in the calculation of matrix elements, as it often involves multicenter integrals, which has been greatly improved in recent years. Currently, the tight-binding approximation (or LCAO) method has become an effective tool for the quantitative calculation of the properties of insulators, compounds, and some semiconductors.

## 4.2 Second quantization representation of the tight-binding approximation Hamiltonian

In narrow-band problems, electrons primarily exhibit localized orbital motions around the lattice points, with a very small probability of delocalized electron motion from one lattice point to another. In such system,, the tight-binding approximation method is particulary useful for band calculations. Furthermore,, the tight-binding approximation method is not only applicable to single-electron problems but is also an effective tool for addressing many-body problems related to narrow bands. For example, based on the tight-binding approximation, the interaction between electrons can be taken into account to discuss the band magnetism of transition metal d electrons. When considering the effect of lattice vibrations, the problem involves the motion of polarons in ionic crystals,which is a key topic in solid-state theory.

To discuss many-body problems, the tight-binding approximation Hamiltonian in the Wannier representation should be formulated. As a starting point,, this section will focus on the non-interacting electron system in a rigid lattice.For simplicity, the discussion is limited to single-band problems where the energy spectrum is spin-independent.. At this stage, both the band index $n$ and spin index $\sigma$ can be omitted. The state vector formula $\psi(r)=\sum_{n,l}C_{nl}a_n(r-l)$ can be simply denoted as:

$$\psi(r)=\sum_{l}C_{l}a(r-l)$$

(37）

for the interaction between a rigid lattice (i.e. without lattice vibration) and electrons, the system can be described by a periodic potential $V(r)$, and the single-electron Hamiltonian of the system is given by:

$$h(r) = -\frac{\hbar^2}{2m}\nabla^2 + V(r)$$

(38)

following the standard method, the second quantization Hamiltonian of the system is expressed as:

$$H = \int \psi^+(r) h(r) \psi(r) dr$$
$$= \sum_{l,l'} C_l^+ C_{l'} \int a^*(r-l) h(r) a(r-l') dr$$

(39)

According to the tight-binding approximation, the Wannier functions in the above equation are replaced by atomic orbital functions, and only the terms $l' = l$ and $l' = l + \rho$ ($\rho$ is the potential vector between the nearest neighbor lattice points) are considered. This yields the tight-binding approximation expression of *H*:

$$H = \varepsilon(0)\sum_l \hat{n}_l - J\sum_l \sum_\rho C_l^+ C_{l+\rho}$$

(40)

where

$$\hat{n}_l \equiv C_l^+ C_l$$

(41)

represents the electron number operator in the Wannier representation, $\varepsilon(0)$ is the energy of the localized orbital electrons, and *J* is the nearest neighbor overlap integral. Equation (40) is non-diagonal, but it can be easily diagonalized using the transformation relationship between the Wannier and Bloch representations. In the literature, two equivalent transformation methods are commonly used.

The most direct method involves applying the operator transformation formula between the two representations:

$$C_l = \frac{1}{\sqrt{N}}\sum_k e^{ik\cdot l} C_k$$

(42)

the orthogonal relation a of irreducible representation of $\sum_{R_l} \exp\left[-i(k-k')\cdot R_l\right] = N\delta_{kk'}$ translational group is obtained:

$$\left.\begin{aligned}\sum_l C_l^+ C_l &= \sum_k C_k^+ C_k \\ \sum_l C_l^+ C_{l+\rho} &= \sum_k e^{ik\cdot\rho} C_k^+ C_k\end{aligned}\right\}$$

(43)

substitute equation (40) to diagonalize the TBA Hamiltonian:

$$H = \sum_k \left\{ \varepsilon(0) - J\sum_\rho e^{ik\cdot\rho} \right\} C_k^+ C_k = \sum_k E(k)\hat{n}_k$$

(44)

and the tight-binding approximation energy band curve can be obtained:

$$E(k) = \varepsilon(0) - J\sum_\rho e^{ik\cdot\rho}$$

in equation (40), $\hat{n}_k \equiv C_k^+ C_k$ represents the electron number operator of the Bloch state.

Alternatively, we can also use the transformation relationship between Bloch state $|n_k\rangle$ and wanier state $|n_l\rangle$ to directly average $H$ to get $E(k)$, which is another way. According to formula $\psi_{nk}(r) = N^{-1/2}\sum_l e^{ik\cdot l} a_n(r,l)$, the transformation relationship of States is given by:

$$|n_k\rangle = \frac{1}{\sqrt{N}}\sum_l e^{ik\cdot l}|n_l\rangle$$

(45)

where $n_k$ is the occupation number of Bloch electrons in the $k$ state, and $n_l$ is the occupation number of electrons in the orbital localized state around the $l$ lattice point. Under the single electron approximation, the diagonal average of Bloch states for $H$ can be obtained through direct computation.

$$\begin{aligned}E(k) &= <1_k|H|1_k> \\ &= \varepsilon(0)\left\{\frac{1}{N}\sum_l <1_k|C_l^+ C_l|1_k>\right\} - J\left\{\frac{1}{N}\sum_l\sum_\rho e^{ik\cdot\rho} <1_l|C_l^+ C_{l+\rho}|1_{l+\rho}>\right\} \\ &= \varepsilon(0) - J\sum_\rho e^{ik\cdot\rho}\end{aligned}$$

(46)

This approach yields the same result as the previous method. However, the latter scheme is more convenient for extending the discussion on the influence of electron-phonon interactions on the bandwidth in narrow-band systems.

## 5. Hamiltonian

For BC model，we consider the positive charge background (b) and the electron (e) as a system, and write their Hamiltonian sums and their interactions, respectively. Aside from the electron kinetic energy, only electrostatic coulomb interactions are taken into account. The Hamiltonian is expressed as:

$$H = H_b + H_e + H_{eb} \tag{47}$$

$$H_e = \sum_{i=1}^{N} \frac{P_i^2}{2m} + \frac{1}{2} e_1^2 \sum_{i=1}^{N} \sum_{\substack{j=1 \\ i \neq j}}^{N} \frac{1}{\left| r_i - r_j \right|} e^{-\mu \left| r_i - r_j \right|} \tag{48}$$

$$H_b = \frac{1}{2} e_1^2 \int \mathrm{d}^3 x \int \mathrm{d}^3 x' \frac{n(x) n(x')}{\left| x - x' \right|} e^{-\mu \left| x - x' \right|} \tag{49}$$

$$H_{eb} = -e_1^2 \sum_{i=1}^{N} \int \mathrm{d}^3 x \frac{n(x)}{\left| x - r_i \right|} e^{-\mu \left| x - r_i \right|} = -e_1^2 \sum_{i=1}^{N} \frac{N}{V} 4\pi \int \mathrm{d}z \frac{e^{-\mu z}}{z} = -4\pi e_1^2 \frac{N^2}{V \mu^2} \tag{50}$$

The shielding factor $\mu$ is introduced to the equation to account for screening effects. $r_i$ represents the *ith* electronic position. $x$ represents the background position. $e$ is the basic charge, $e_1 = e/\sqrt{4\pi\varepsilon_0}$ . Since the coulomb force is a long-range force，while $N \to \infty$， $V \to \infty$ ， $N/V$ remains unchanged，if the coulomb potential energy is calculated separately， $H_e$ 、 $H_b$ and $H_{eb}$ will diverge. Physically, when increasing the volume by keeping the electron density ( $N/V$ ) constant, the average electron energy $H/N$ should be a constant. Therefore, the divergent terms in the $H_e$ , $H_b$ and $H_{eb}$ should be able to cancel each other out，in order to see this, We have added an attenuation factor $\mu$ to (48), (49), and (50)，after the integration $L$ of the cube with a side length of $L \to \infty$ ， And then order $\mu \to 0$，this is a commonly used technique in solid state physics calculations. $e_1 n(x)$ is the charge density at background $x$ , and $n(x) = N/V$ is a constant.

$$H_b = \frac{1}{2} e_1^2 \left( \frac{N}{V} \right)^2 \int \mathrm{d}^3 x 4\pi \int \mathrm{d}z \frac{e^{-\mu z}}{z} = 4\pi e_1^2 \frac{N^2}{2V\mu^2}$$

（51）

$$H_{eb} = -e_1^2 \sum_{i=1}^{N} \frac{N}{V} 4\pi \int \mathrm{d}z \frac{e^{-\mu z}}{z} = -4\pi e_1^2 \frac{N^2}{V\mu^2}$$

（52）

$$\sum_{i=1}^{N} \frac{P^2}{2m} = \sum_{l'\sigma'} \sum_{l\sigma} a^{\dagger}_{k_{l'}\sigma'} \left\langle k_{l'}\sigma' \left| \frac{P^2}{2m} \right| k_l \sigma \right\rangle a_{k_l\sigma} = \sum_{l\sigma} \frac{\hbar^2 k_l^2}{2m} a^{\dagger}_{k_l\sigma} a_{k_l\sigma}$$

(53)

In the above formula, use $k_l$ representative $k_{xl}$ 、 $k_{yl}$ and $k_{zl}$ ， $l = 0, \pm1, \pm2, \pm3, \cdots$ for typographical convenience， we sometimes use $k$ and $k'$ in lieu of $k_l$ and $k_{l'}$， just remember the momentum $k$ take discrete values instead of continuous values. The second item in（48） is

$$\frac{1}{2} e_1^2 \sum_{l'\sigma'} \sum_{m'\lambda'} \sum_{l\sigma} \sum_{m\lambda} a^{\dagger}_{k_{l'}\sigma'} a^{\dagger}_{k_{m'}\lambda'} (k_{l'}\sigma', k_{m'}\lambda' \left| \frac{e^{-\mu|r_1 - r_2|}}{|r_1 - r_2|} \right| k_l\sigma, k_m\lambda) a_{k_m\lambda} a_{k_l\sigma}$$

$$= \frac{1}{2} e_1^2 \sum_{l'} \sum_{m'} \sum_{l\sigma} \sum_{m\lambda} \delta_{\sigma\sigma'} \delta_{\lambda\lambda'} a^{\dagger}_{k_{l'}} a^{\dagger}_{k_{m'}} (k_{l'} k_{m'} \left| \frac{e^{-\mu|r_1 - r_2|}}{|r_1 - r_2|} \right| k_l k_m) a_{k_m} a_{k_l}$$

（54）

The matrix elements are first calculated in the position representation. Insert two completeness relations in the matrix element in the above equation:

$$\int d^3 r^a \int d^3 r^\beta \left| r^\alpha r^\beta \right\rangle \left\langle r^\alpha r^\beta \right| = 1$$

$$\int d^3 r^\gamma \int d^3 r^\delta \left| r^\gamma r^\delta \right\rangle \left\langle r^\gamma r^\delta \right| = 1$$

And

$$(k_{l'} k_{m'} \left| \frac{e^{-\mu|r_1 - r_2|}}{|r_1 - r_2|} \right| k_l k_m) = \frac{1}{V^3} \int d^3 r^a \int d^3 r^\beta e^{-i(k_{l'} - k_l)\cdot r^a} e^{-i(k_{m'} - k_m)\cdot r^\beta} \frac{e^{-\mu|r^\alpha - r^\beta|}}{|r^\alpha - r^\beta|}$$

$$= \delta\left(k_{l'} + k_{m'}, k_l + k_m\right) \frac{1}{V} \int \frac{e^{-\mu r}}{r} e^{-i(k_{l'} - k_l)\cdot r} d^3 r$$

$$= \delta\left(k_{l'} + k_{m'}, k_l + k_m\right) \frac{2\pi}{V} \int r e^{-\mu r} \left( \frac{2}{kr} \sin kr \right) dr$$

$$= \delta\left(k_{l'} + k_{m'}, k_l + k_m\right) \frac{1}{V} \frac{4\pi}{\left(k_{l'} - k_l\right)^2 + \mu^2}$$

（55）

In the above calculation, the second step uses the independent variable substitution：

$r^{\alpha}+r^{\beta}=r'$ 、 $r^{\alpha}-r^{\beta}=r$ [Jacobian determinant $J=\left(1/2\right)^{3}$]， substitute this equation back (54)， $k_l=k$ ， $k_m=k'$， $k_{l'}=k+q$ ， $k_{m'}=k'-q$ ， the second item in $H_e$ is

$$\frac{e_1^2}{2V}\sum_k\sum_{k'}\sum_q\sum_{\sigma\lambda}\frac{4\pi}{q^2+\mu^2}a_{k+q,\sigma}^{\dagger}a_{k'-q,\lambda}^{\dagger}a_{k'\lambda}a_{k\sigma}$$

（56）

Separate out $q=0$ is

$$\frac{e_1^2}{2V}\sum_k\sum_{k'}\sum_q\sum_{\sigma\lambda}\frac{4\pi}{\mu^2}a_{k\sigma}^{\dagger}a_{k'\lambda}^{\dagger}a_{k'\lambda}a_{k\sigma}=\frac{e_1^2}{2V}\frac{4\pi}{\mu^2}\left(N^2-N\right)$$

$N$ is the total number of particles， because

$$a_{k\sigma}^{\dagger}a_{k'\lambda}^{\dagger}a_{k'\lambda}a_{k\sigma}=N_{k\sigma}N_{k'\lambda}-N_{k\sigma}\delta_{kk'}\delta_{\sigma\lambda}$$

Now substitute (53) and (56) back to equation (48), $H_e$ is

$$H_e=\sum_{k\sigma}\frac{\hbar^2k^2}{2m}a_{k\sigma}^{\dagger}a_{k\sigma}+\frac{e_1^2}{2V}\sum_q{}^{*}\sum_{k\sigma}\sum_{k'\lambda}\frac{4\pi}{q^2+\mu^2}a_{k+q,\sigma}^{\dagger}a_{k'-q,\lambda}^{\dagger}a_{k'\lambda}a_{k\sigma}+\frac{e_1^2}{2V}\frac{4\pi}{\mu^2}\left(N^2-N\right)$$

（57）

In the formula, the $q$ on the sum sign of "*" indicates that the part where q=0 is ignored during the sum. When $V\to\infty, N\to\infty$, while keeping *N*/*V* constant, the last term to the right of the equal sign causes the average energy $H_e/N$ of each particle to become:

$$\frac{1}{2}4\pi e_1^2\left(\frac{N}{V}\right)\frac{1}{\mu^2}-\frac{1}{2}4\pi e_1^2\left(\frac{N}{V}\right)\frac{1}{N}\frac{1}{\mu^2}$$

The first term is constant, and the second term approaches zero. However, if $\mu\to 0$, the first term becomes a divergent term. However, this term just cancels out the divergent $H_b$ and $H_{eb}$ term. Thus, the Hamiltonian of the system becomes

$$H=\sum_{k\sigma}\frac{\hbar^2k^2}{2m}a_{k\sigma}^{\dagger}a_{k\sigma}+\frac{e_1^2}{2V}\sum_q\sum_{k'}\sum_{\sigma\lambda}\frac{4\pi}{q^2}a_{k+q,\sigma}^{\dagger}a_{k'-q,\lambda}^{\dagger}a_{k'\lambda}a_{k\sigma}=H_0+H_1$$

(58)

**6. Energy level**

## 6.1 Establishment of the Hamiltonian operator

In a multi electron atom, the kinetic energy operator of the *i*-th electron is $-\frac{\hbar^2}{2m_e}\nabla_i^2$, its potential energy in the average potential field $V(r_i)$ generated by the nucleus and other $(Z-1)$ electrons is $V(r_i)$. Then the Hamiltonian operator of the whole atom is $\hat{H}=\sum_{i=1}^{Z}\left[-\frac{\hbar^2}{2m_e}\nabla_i^2+V(r_i)\right]$.

## 6.2 Separating variables and single electron equation

Let the wave function $\Psi\left(\vec{r}_1,\vec{r}_2,\cdots,\vec{r}_Z\right)=\psi_1\left(\vec{r}_1\right)\psi_2\left(\vec{r}_2\right)\cdots\psi_Z\left(\vec{r}_Z\right)$ be substituted into the time-dependent Schrodinger equation $i\hbar\frac{\partial\Psi}{\partial t}=\hat{H}\Psi$ to obtain

$$i\hbar\frac{\partial\Psi}{\partial t}=\sum_{i=1}^{Z}\left[-\frac{\hbar^2}{2m_e}\nabla_i^2+V(r_i)\right]\Psi \tag{59}$$

Since $\Psi\left(\vec{r}_1,\vec{r}_2,\cdots,\vec{r}_Z\right)=\psi_1\left(\vec{r}_1\right)\psi_2\left(\vec{r}_2\right)\cdots\psi_Z\left(\vec{r}_Z\right)$, then

$$i\hbar\frac{\partial\Psi}{\partial t}=\sum_{i=1}^{Z}\left[-\frac{\hbar^2}{2m_e}\nabla_i^2+V(r_i)\right]\psi_1\left(\vec{r}_1\right)\psi_2\left(\vec{r}_2\right)\cdots\psi_Z\left(\vec{r}_Z\right) \tag{60}$$

After separating the variables, the steady-state Schrödinger equation of the single electron is obtained as

$$\left[-\frac{\hbar^2}{2m_e}\nabla_i^2+V(r_i)\right]\psi_i\left(\vec{r}_i\right)=E_i\psi_i\left(\vec{r}_i\right) \tag{61}$$

## 6.3 Equation representation in spherical coordinate system

For spherically symmetric potential energy $V(r_i)$, $\psi_i\left(\vec{r}_i\right)$ is represented as $\psi_i\left(\vec{r}_i\right)=R_{nl}(r)Y_{lm}(\theta,\varphi)$ in the spherical coordinate system.

The single electron steady state Schrödinger equation in spherical coordinates is

$$\left[-\frac{\hbar^2}{2m_e}\left(\frac{1}{r^2}\frac{\partial}{\partial r}\left(r^2\frac{\partial}{\partial r}\right)+\frac{1}{r^2\sin\theta}\frac{\partial}{\partial\theta}\left(\sin\theta\frac{\partial}{\partial\theta}\right)+\frac{1}{r^2\sin^2\theta}\frac{\partial^2}{\partial\varphi^2}\right)+V(r)\right]\psi_i\left(\vec{r}_i\right)=E_i\psi_i\left(\vec{r}_i\right)$$

(62)

Substitute $\psi_i\left(\vec{r}_i\right)=R_{nl}(r)Y_{lm}(\theta,\varphi)$ into **Eq (62)**, and separate the variables to obtain the angular equation and radial equation as

$$\left[-\frac{\hbar^2}{2m_e}\frac{1}{r^2}\frac{d}{dr}\left(r^2\frac{dR_{nl}(r)}{dr}\right)+V(r)R_{nl}(r)\right]+\frac{l(l+1)\hbar^2}{2m_er^2}R_{nl}(r)=E_{nl}R_{nl}(r)$$

(63)

The solution of the angular equation is spherical harmonic function $Y_{lm}(\theta,\varphi)$, which is determined by the quantum numbers $l$ and $m$.

## 6.4 Introduction of effective potential energy and solution

**To solve the radial equation, the effective potential energy**

$V_{eff}(r)=V(r)+\frac{l(l+1)\hbar^2}{2m_er^2}$ is introduced, and the radial equation takes the form:

$$-\frac{\hbar^2}{2m_e}\frac{1}{r^2}\frac{d}{dr}\left(r^2\frac{dR_{nl}(r)}{dr}\right)+V_{eff}(r)R_{nl}(r)=E_{nl}R_{nl}(r)$$

(64)

For the solution of **Eq (64)**, combined with the boundary conditions, when $r\to 0$, $R_{nl}(r)$ is finite; When $r\to\infty$, $R_{nl}(r)\to 0$, the expression of the energy eigenvalue $E_{nl}$ related to the principal quantum number $n$ and the angular quantum number $l$ can be obtained.

## 6.5 Formula for determining effective nuclear charge number and energy level

Using the central force field approximation, $V(r)$ is approximated to

$V(r)=-\frac{Z^*e^2}{4\pi\varepsilon_0 r}$, where $Z^*=Z-\sigma$ and $\sigma$ are shielding constants, which can be determined by the slay rule, etc.

By substituting $V(r) = -\frac{Z^* e^2}{4\pi\varepsilon_0 r}$ into the expression of energy eigenvalue obtained by solving the radial equation **Eq (64)**, the energy level formula $E_{n,l} = -\frac{m_e (Z^* e^2)^2}{8\varepsilon_0^2 \hbar^2 n^2} = -\frac{R_H Z^{*2}}{n^2}$ of multi electron atom can be obtained, where $R_H = \frac{m_e e^4}{8\varepsilon_0^2 \hbar^2}$ is the Rydberg constant.